\documentclass[iop]{emulateapj}
\usepackage{txfonts}
\usepackage{graphicx}
\usepackage{rotating}
\usepackage{lscape}
\usepackage{hyperref}
\usepackage{color,hyperref}
\definecolor{linkcolor}{rgb}{0,0,0.5}
\definecolor{notecolor}{rgb}{0.8,0,0}
\hypersetup{colorlinks=true, linkcolor=linkcolor, citecolor=linkcolor,
  filecolor=linkcolor, urlcolor=linkcolor}

\def\cor{$\alpha=18^{\mathrm{h}}33^{\mathrm{m}}43.94^{\mathrm{s}}, \delta=-07\arcdeg36\arcmin52.8\arcsec$}
\def\co{$^{12}$CO (1-0)}
\def\13co{$^{13}$CO (1-0)}
\def\c18o{C$^{18}$O (1-0)}
\def\msun{$M_{\odot}$}

\def\cm2{cm$^{-2}$}
\def\vel{km s$^{-1}$}
\def\g24{G24.136}
\def\um{$\mu$m}
\journalinfo{The Astrophysical Journal, \textrm{797:1(13pp), 2014 December 10}}
\submitted{Received 2014 March 20; accepted 2014 October 22}

\shorttitle{Triggered star formation around G24.136+0.436}

\begin{document}

\title{A feedback-driven bubble G24.136+00.436: a possible site of triggered star formation}

\author{Hong-Li Liu\altaffilmark{1,2$\dag$}, Yuefang Wu\altaffilmark{3$\star$}, JinZeng Li\altaffilmark{1}, Jing-Hua Yuan\altaffilmark{1},Tie Liu\altaffilmark{4},Xiaoyi Dong\altaffilmark{3}}

\altaffiltext{1}{National Astronomical Observatories, Chinese Academy of Sciences, 20A Datun Road, Chaoyang District, Beijing 100012, China}
\altaffiltext{2}{University of Chinese Academy of Sciences, 100049 Beijing, China}
\altaffiltext{3}{Department of Astronomy, Peking University, 100871 Beijing, China}
\altaffiltext{4}{Korea Astronomy and Space Science Institute 776, Daedeokdae-ro, Yuseong-gu, Daejeon, Republic of Korea 305-348}
\altaffiltext{$\dag$}{hlliu@nao.cas.cn}
\altaffiltext{$\star$}{yfwu.pku@gmail.com}

\begin{abstract}
We present a multi-wavelength study of the IR bubble G24.136+00.436. The J=1-0 observations of $^{12}$CO, $^{13}$CO and C$^{18}$O
were carried out with the Purple Mountain Observatory 13.7 m telescope.  Molecular gas with a velocity of 94.8 \vel\
is found prominently in the southeast of the bubble, shaping as a shell with a total mass of $\sim2\times10^{4}$ \msun.
It is likely assembled during the expansion of the bubble. The expanding shell consists of six dense cores.
Their dense (a few of $10^{3}$ cm$^{-3}$) and massive (a few of $10^{3}$ \msun) characteristics coupled with
the broad linewidths ($>$ 2.5 \vel) suggest they are promising sites of forming high-mass stars or clusters.
This could be further consolidated by the detection of compact HII regions in Cores A and E.
We tentatively identified and classified 63 candidate YSOs based on the \emph{Spitzer} and UKIDSS data.
They are found to be dominantly distributed in regions with strong emission of molecular gas, indicative of active
star formation especially in the shell. The HII region inside the bubble is mainly ionized by a $\sim$O8V star(s),
of the dynamical age $\sim$1.6 Myr. The enhanced number of candidate YSOs and secondary star formation in the shell
as well as time scales involved, indicate a possible scenario of triggering star formation, signified by
the ``collect and collapse" process.

\end{abstract}


\section{Introduction}\label{s1}
It is believed that the majority of stars in Galaxy form in clusters. The feedback from them is expected to have
strong influences on their neighbor interstellar mediums (ISMs). Massive stars residing in a cluster deposit large
amounts of energies into ambient molecular clouds in the form of strong ultraviolet (UV) radiation (h$\nu >$ 13.6 ev)
which photoionizes surrounding gas, creating bubbles/HII region. The expanding bubbles/HII regions likely prompts the
collapse of  molecular clouds which may not contract and fragment spontaneously, and stimulates star formation of next
generations. This process is defined as triggered star formation.

Scenarios of triggered star formation have been suggested by a few theoretical models and several observations.
For example, \citet{elm77} first proposed a ``collect and collapse"  process. It could be described as follows:
as an HII region expands outwards, its surroundings may be compressed and collected between the ionization
front (IF) and the shock front (SF); the shell between the IF and SF in due time may become dense, massive
and gravitationally unstable, and collapse to form new stars. If this process occurs, it would self-propagate and
lead to sequential star formation. Several simulations concluded that the expanding HII region or ionizing
feedback should be an efficient trigger for star formation in molecular clouds if the mass of the ambient molecular
material is massive enough \citep{hos05,hos06,dal07,dal12,dal13}.  Observational signatures of this process have
been presented on boundaries of several HII regions where the number of young stellar objects (YSOs) is enhanced
(e.g. \citet{deh05, pom09,liu12}).

 Other existing observations also showed the prevalence of triggering star formation in the Milky Way.
For example, \citet{tho12} studied the association of Red MSX massive young stellar objects (MYSOs) with the
322 bubbles from the \citet{chu06} catalog. The result suggested that around $14\% - 30\%$ of massive stars
in the Milky Way might have been triggered by the expansion of the bubbles. Similarly, from  the association
of MYSOs with $1018$ Milky Way Project bubbles (visually identified by citizen scientists \citep{sim12}),
\citet{ken12} found that approximately $22\%\pm2\%$ of MYSOs might have been induced by feedback from the
expanding bubbles/HII regions. Additionally, by combining the geometry of stars and surviving cold ISMs,
some possible signatures of triggered star formation have been reported over a number of known HII regions,
such as Sh2$-212$ \citep{deh08}, W51a \citep{kan09}, Sh2$-217$ \citep{bra11}, and Sh2$-90$ \citep{sam14}.
These results suggested that IR bubbles or HII regions could serve as a good laboratory for the study of
triggered star formation. However, the exact processes of interactions between bubbles/HII regions and the
ambient ISMs remain unclear. Hopefully, the investigation of the physical connection and interaction of
bubbles/HII regions with their adjacent ISMs would achieve a better understanding of star formation around
bubbles/HII regions \citep{sam14}.

\begin{figure}
  \begin{center}
 \centering
  \includegraphics[width=0.48\textwidth]{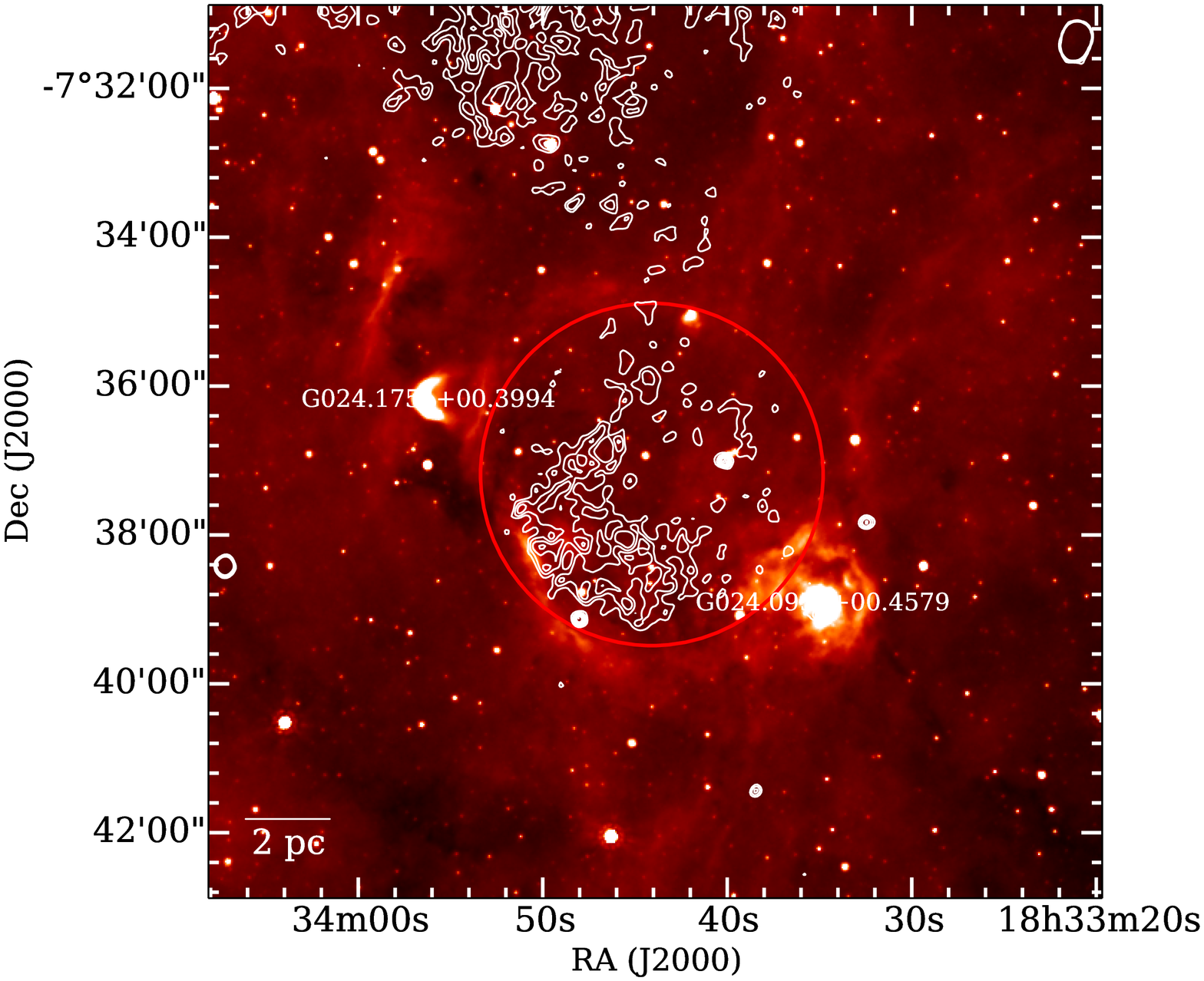}
  \caption{8.0 \um\ image of \g24\ overlaid with contours of free-free emission at 20 cm. The contours start from $5\sigma$ ($\sigma=0.15$ mJy/beam)
  in a step of $1\sigma$.
  The image with a size of $12\arcmin \times 12 \arcmin$ is centered at \cor. The circle characterizes
  the radius of the bubble, 2.3$\arcmin$, which corresponds to 3.9 pc at a distance of 5.6 kpc. A scale bar of
  2 pc is shown on the bottom left.}
  \label{micron8}
  \end{center}
\end{figure}

The purpose of this work is to probe the evidence of triggered star formation and investigate formation of massive
stars or clusters as well as star formation histories around the bubbles. Therefore, we carried out a multi-wavelength
study towards the bubble MWP1G024.136+00.436 (\citet{sim12}, G24.136 hereafter), located at
$18^{\mathrm{h}}33^{\mathrm{m}}43.96^{\mathrm{s}}, \delta=-07\arcdeg37\arcmin15.5\arcsec$. This paper is organized
as follows: the region of G24.136 is presented in Sect. 2; the observations of J=1-0 transitions of \co, \13co and \c18o
and  the archival infrared and radio data are described in Sect. 3; the result is provided in Sect. 4; the discussion is
arranged in Sect. 5; the last section is the summary.

\section{Presentation of \g24}\label{sg24}
 In Fig. \ref{micron8}, the bubble \g24\ appears to be a complete ring (red circle) as seen in the 8 \um\ image.
The 8 \um\ emission predominantly comes from two prominent polycyclic aromatic hydrocarbons (PAHs) at 7.7 and 8.6 \um,
which are indicative of photodissociation regions (PDRs). Since PDRs are possible nurseries of massive star or cluster
formation \citep{zav10}, the bright ring consisting of PDRs demonstrates possibly rich activities of star formation in \g24.
The free-free continuum emission from ionized regions almost fills the bubble. This picture is greatly consistent with
the fact that the bubble shrouds the HII region G024.139+0.427 \citep{loc96}. Additionally, the region in the northeast
to \g24\ is also full of the free-free radiation, indicating another HII region. We are not sure whether this region is
located at the same distance as the bubble due to lacking more information (e.g., kinetic information), therefore,
exclude it from further analysis. In the edge of \g24\ two compact bright PDRs are consistent with conspicuously
strong free-free emission. One of the compact sources located at the southwest, G024.0953+00.4579, is classified as
an HII region by \citet{urq09} with the 6 cm continuum VLA survey. The other compact source G024.1754+00.3994 is
located at the northeast . In Sect. \ref{sCHIIregion} we will confirm that these two regions are indeed compact HII
(CHII) regions. The classical HII region G024.139+0.427 with its two neighboring CHII regions shows a multi-generation
of star formations. There are also some regions in the rim of \g24\ seen as absorption in the 8 \um\ image. The associated
\co\ and \13co\ emission (Fig. \ref{inten}) suggests they lie behind the cold gas.

 Adopting the \citet{bra93} rotation curve, the systemic velocity, $V_{lsr} = 94.1$ \vel, yielded a distance
of $5.2 \pm 1.0$ kpc \citep{uqr11b}. Similarly, assuming the \citet{mcc07} rotation curve, the systemic velocity,
$V_{lsr} = 98.4$ \vel\ , resulted a distance of $5.9 \pm 1.0$ kpc \citep{and09}. In this work, we adopt the mean value
of both distances, $5.6 \pm 1.0$ kpc. Therefore, a radius of 2.3$^\prime$ \citep{sim12} of \g24\ corresponds to $3.9 \pm 0.7$ pc.

\section{Observations and Archival Data}\label{s2}
\subsection{Millimeter Line Observations}
The observations of J=1-0 of $^{12}$CO, $^{13}$CO and C$^{18}$O were made in May, 2013 using the Purple Mountain
Observatory 13.7m telescope in Delingha, China. The telescope was configured with an SIS superconducting receiver
consisting of a nine-beam array in the front end. $^{12}$CO and its isotopologues were observed simultaneously
with excellent positional registrations and calibrated at the upper sideband (USB) and the lower sideband (LSB)
respectively. Pointing and tracking accuracies were better than 5$\arcsec$. The backend was composed of an FFT
spectrometer with 1 GHz bandwidth and 16384 channels. The corresponding velocity resolutions are 0.16 km s$^{-1}$
for $^{12}$CO (1-0), 0.17 km s$^{-1}$ for $^{13}$CO (1-0) and C$^{18}$O (1-0) lines, respectively. The half-power
beamwidth was about 54$\arcsec$ and the main beam efficiencies at observed frequencies were 0.44 (115.271202 GHz)
and 0.48 ((110.201353 GHz) respectively. The On-The-Fly (OTF) mapping mode was adopted. Maps were made with the
beam center running on a 12$\arcmin$ $\times$ 12$\arcmin$ region centered at \cor. The scan rate was 50$\arcsec$
per second and the integrated time was 0.3 s. To improve the ratio of signal to noise, six observations were
performed on the same region. During the observations, the system temperatures ranged from 370 to 430 K for
the USB and from 250 to 270 K for the LSB. After observations, all OTF data were merged into a data cube with
a grid spacing of 30$\arcsec$. The antenna temperatures were converted into main-beam brightness temperatures.
The RMS noise levels of the resulting data were $\sim$0.6 K for $^{12}$CO (1-0), and $\sim$0.3 K for $^{13}$CO
(1-0) and C$^{18}$O (1-0) lines. The resulting data were analyzed and visualized with the software GILDAS \citep{gui00}.

\subsection{Archival Data}
The GLIMPSE survey \citep{ben03} had mapped portions of the inner Galactic plane, using Infrared Array Camera
(IRAC) \citep{faz04} aboard the \emph{Spitzer} Space Telescope. From the GLIMPSE Spring '07 Archive, 10509 point
sources over the region of $12\arcmin \times 12\arcmin$ were retrieved, which are complete to a magnitude of
14.3 mag at 3.6 \um, 14.9 mag at 4.5 \um, 12.6 mag at 5.6 \um\ and 12.0 mag at 8.0 \um, respectively.
The images of four IRAC bands of \g24\ were also used to reveal its physical characteristics.
 The 24 $\mu$m image of \g24\ was obtained from MIPSGAL, another survey of the inner Galactic plane using
the Multiband Imaging Photometer for \emph{Spitzer} (MIPS) instrument \citep{rie04}. The point sources were
extracted from the image using the DAOPHOT package of software IRAF (the detailed descriptions of photometry
are prepared in later work (Liu et al. 2014, in preparation)). We found 63 MIPS objects match the IRAC sources
within a $2\arcsec$ search radius. These MIPS sources are completed to 8.5 mag at 24 \um. Additionally,
deep near-infrared (NIR) point sources were retrieved from the UKIDSS 6th data release of Galactic Plane Survey
(GPS) \citep{luc08,law07}. There were 3184 UKIDSS sources matched well with IRAC counterparts within
the $2\arcsec$ radius. These NIR sources are complete to a magnitude of 14.5 mag, 15.0 mag and 14.0 mag in
the $J$, $H$, and $Ks$ band, respectively.

To calculate the Lyman continuum photons from the central region of \g24, we used radio continuum emission at
20 cm from MAGPIS \citep{hel06}. To correct for the deficiency of missing fluxes, MAGPIS combined  VLA images
with those from a 1.4 GHz survey performed with the Effelsberg 100 m telescope \citep{rei90}. As a result, a
better resolution of $5.4\arcsec$ and  1$\sigma$ sensitivity of $<0.15 $ mJy were achieved.

\section{Results}\label{s3}

\subsection{Molecular Emission} \label{s3.2.1}

\begin{figure}
  \begin{center}
 \centering
 \includegraphics[width=0.4\textwidth]{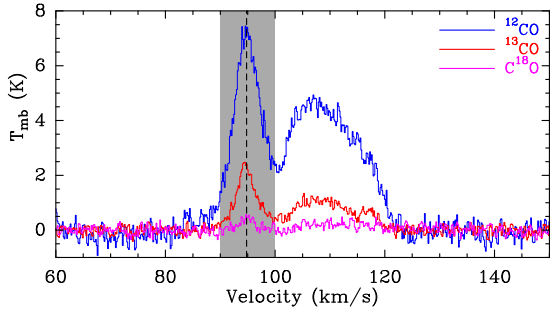}
  \caption{The averaged spectra of $^{12}$CO (1-0), $^{13}$CO (1-0), C$^{18}$O (1-0) over a region of 12 $\arcmin$ $\times$ 12
$\arcmin$ centered at \cor. The black dashed line represents a systemic velocity of 94.8 km s$^{-1}$. The darker shade marks the
narrow-velocity component with an interval from $90$ \vel\ to 100 \vel.}
  \label{average}
  \end{center}
\end{figure}

\begin{figure*}
\centering
\includegraphics[width=0.5\textwidth,angle=270]{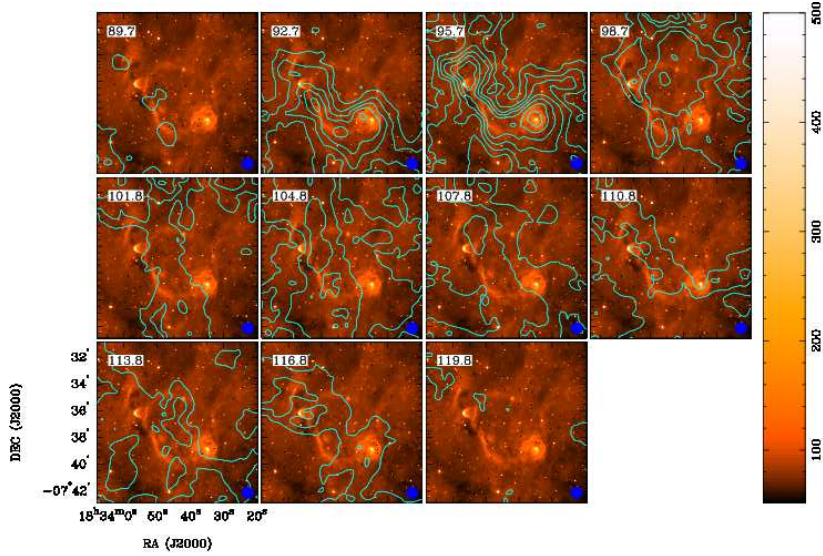}
\caption{The channel map (contours) of \co\ overlaid on the 8 \um\ image (color scale). The contour levels start from $5 \sigma$
($\sigma=0.6$ K \vel) in a step of $3 \sigma$. The velocity of the channel is plotted at the upper-right of each panel. The units of
the 8 \um\ image is MJy/sr. The blue filled circle on the bottom right shows an angular resolution of $54\arcsec$.}
\label{chan12}
\end{figure*}

Figure \ref{average} displays the averaged spectra of $^{12}$CO (1-0), $^{13}$CO (1-0) and C$^{18}$O (1-0) over the region
of 12$\arcmin$ $\times$ 12$\arcmin$. It clearly portrays a narrow-velocity component, accompanying with a broad-velocity
component. Based on the \13co\ average spectrum, their velocities range from $90$ \vel\ to 100 \vel\ and from
$100$ \vel\ to 120 \vel, respectively.

 The narrow-velocity component emission is around a systemic velocity of 94.8 $\pm$ 0.42 km s$^{-1}$ (see Fig. \ref{average}),
from the ambient environments of \g24\ \citep{and09}. The present systemic velocity is well consistent with that derived
from the radio recombination line \citep{loc96}. This supports the association of molecular gas in the interval from
$90$ \vel\ to 100 \vel\ with the bubble. Figure \ref{chan12} present the channel maps of \co\
overlaid on the 8 \um\ image. It is clear that the emission of the narrow-velocity component is enhanced in the ring-like
structure. The velocities along the NW-SE of the bubble show a systemic variance. It is still unclear whether this variance
is due to projecting effects, ``rocket" effects from massive stars, or some others.

 The broad-velocity component appears to be projected on the bubble as an irregular morphology (see Fig. \ref{chan12}).
Its spatial distribution is similar to that of the narrow-velocity component, showing the prominent emission in the
southeastern edge of the bubble. This suggests both components may reveal the same region. However, emission of the
broad-velocity component is much weaker than that of the narrow-velocity component. Figure \ref{average} shows the
non-Gaussian profile of the broad-velocity component. Apart from this, its emission intensities and profiles conspicuously
vary in different positions (see Fig. \ref{core:spec}). These features are similar to those of the supernova remnant (SNRs)
seen by CO and other tracers (e.g., IC443 \citep{den79,wan92,zha10,van93}, W44 \citep{set04} and W28 \citep{ari99}).
\citet{van93} proposed that rapid variations in line profiles from position to position probably indicate the inhomogeneous
composition of the preshocked gas and the gas likely undergoing more than one shock. In this context,
the broad-velocity component of \g24\ may trace shocked gas induced by the expanding bubble.
However, this is a complex region with many star forming regions.
On the one hand, at the Northern side of the bubbles there are three clear velocity peaks between 90 and 120 km s$^{-1}$ instead of
a narrow component along with a broad one. On the other hand, the location of the bubble at $l\sim24\arcdeg$ is roughly towards the Scutum-Centaurus
tangent point near the end of the long bar in the Galaxy. Therefore, the line profile may also be simply a result of blended lines
at different velocities. The lacking of more
information on the broad-velocity component makes the determination of its origin and nature difficult. As a result, we focus
on the narrow velocity component in the following analysis.

\subsection{Dense Molecular Cores}
\begin{figure}
\centering
\includegraphics[width=0.48\textwidth]{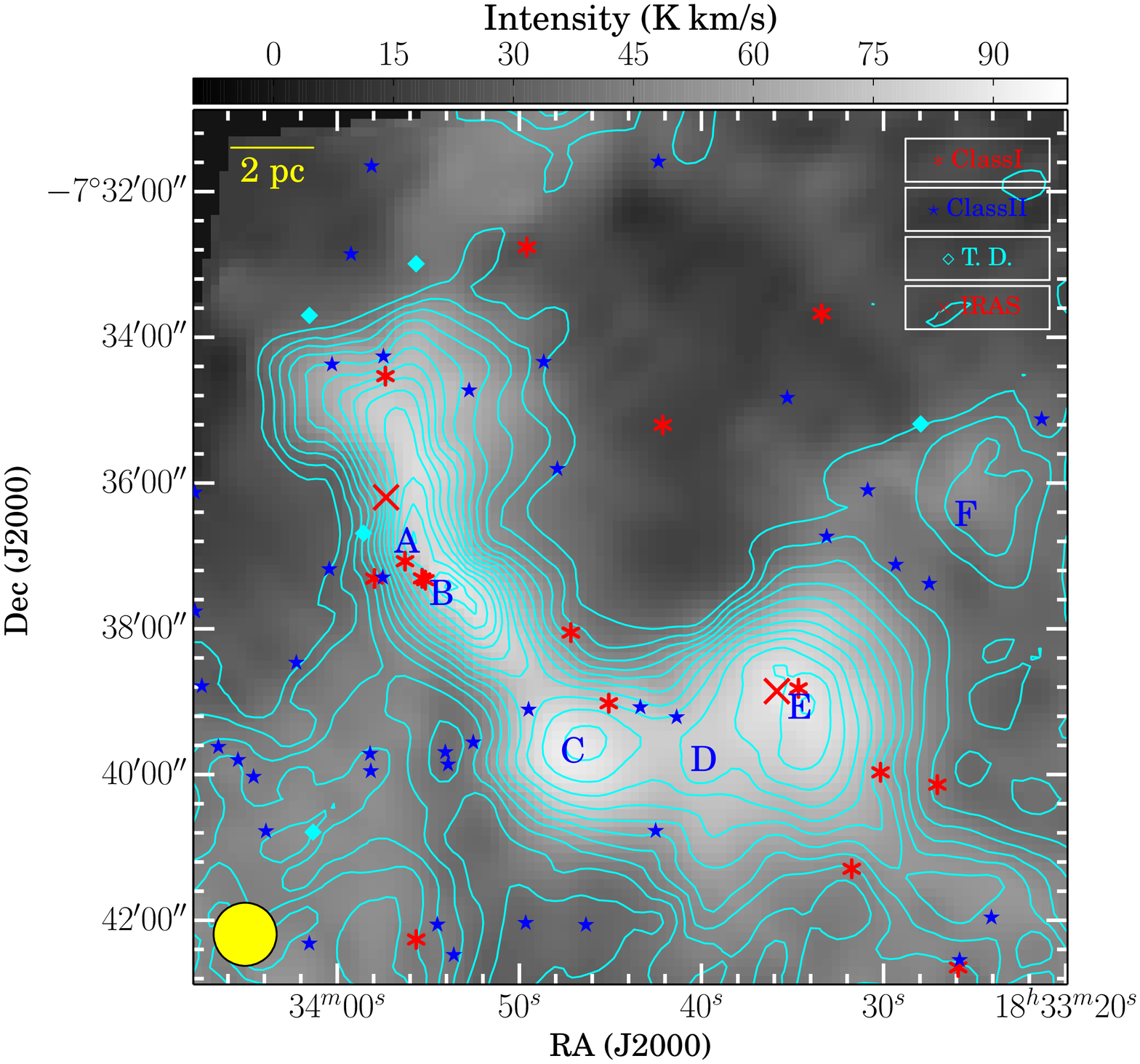}
\caption{$^{13}$CO (1-0) velocity-integrated contours overlaid on the $^{12}$CO (1-0) integrated map (gray scale). The velocity covers 90 km s$^{-1}$ to 100
km s$^{-1}$. Levels of the contour of $^{13}$CO (1-0)
emission start from $10 \sigma$ ($\sigma=0.7$ K \vel) with a step of $3 \sigma$.  Molecular cores are denoted as ``A"
to ``F".  The red asterisks represent Class I YSOs, the blue stars Class II YSOs, the cyan diamonds transition disks sources,
and the red crosses IRAS sources.
The yellow filled circle on the bottom left displays an angular resolution of $\theta_{\mathrm{beam}}=54^{\prime\prime}$. A scale bar of 2 pc
is shown on the top left.}
\label{inten}
\end{figure}
\begin{figure*}
\centering
\includegraphics[width=0.8\textwidth]{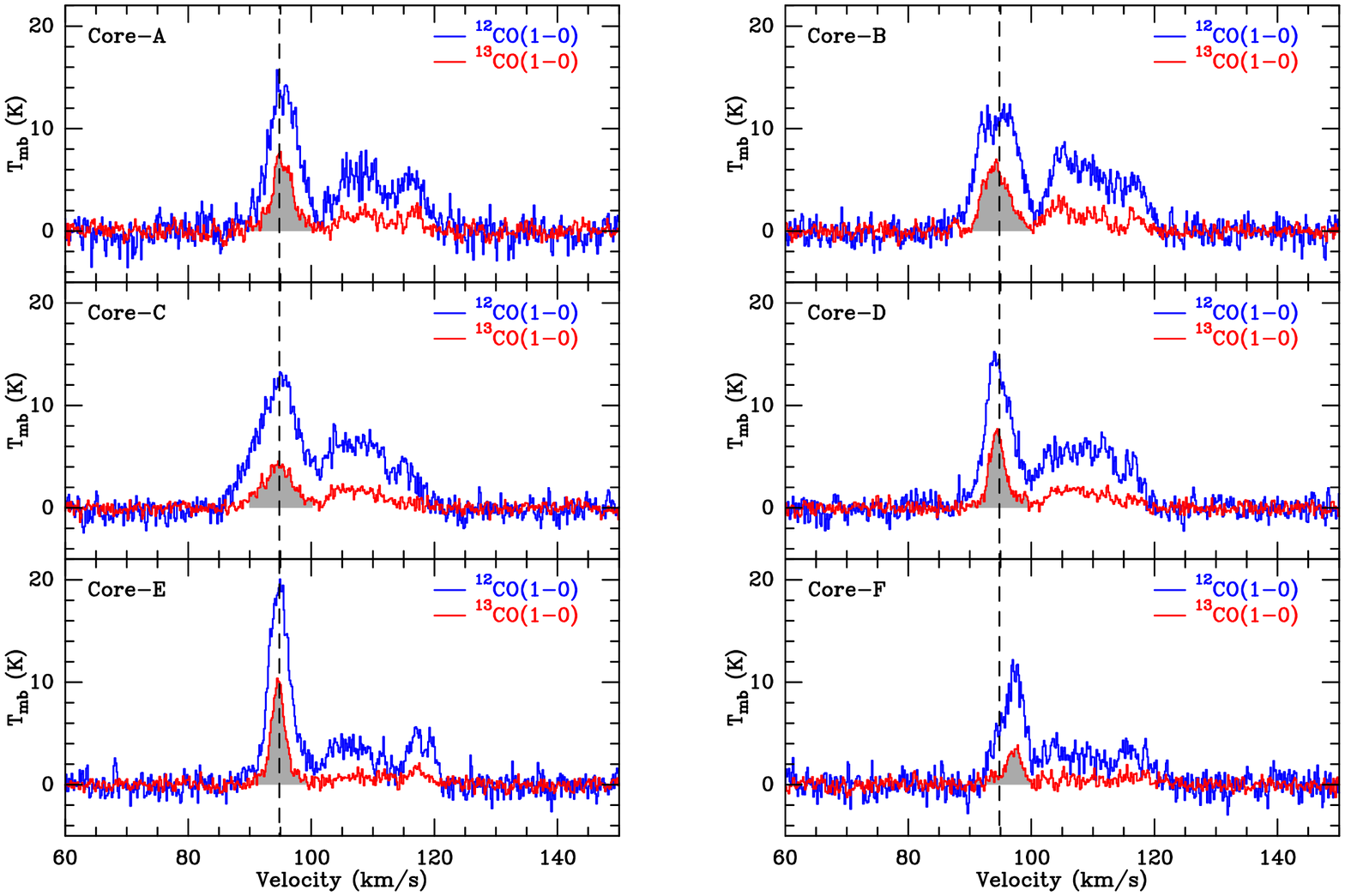}
\caption{The spectra of $^{12}$CO (1-0) and $^{13}$CO (1-0) at the peaks of six cores. The  black dashed lines represent
a systemic velocity of 94.8 km s$^{-1}$. The core names are labeled on the top left of each panel.
The darker shades show the narrow-velocity component with an interval from $90$ \vel\ to 100 \vel.}
\label{core:spec}
\end{figure*}

Figure \ref{inten} presents the \co\  integrated map overlaid with contours of \13co\  intensity. The \co\ emission
evidently shows a molecular ring around the infrared bubble. The south-east part is dense and extensive while the
north-west is diffuse and rarefied. The spatial distribution of \13co\ is not as extended as that of \co,
indicating that the \13co\ transition traces more denser regions than \co. In the southeast edge of the bubble the
dense region sculpted as a shell suggests that it may be gathered during the expansion of the bubble.

 Figure \ref{inten} shows that the expanding shell consists of several substructures, whereby we extracted
them from the map of the \13co\ integrated intensity using the 2D Clumpfind algorithm \citep{wil94}. Clumpfind
searches for local peaks of emission and uses closed contours at lower intensity levels to assign boundaries.
Taking the contour levels used in Fig. \ref{inten} as the inputting parameters, Clumpfind yielded six cores within
the shell. They are marked as ``A",``B",``C",``D",  ``E", ``F" hereafter. Their parameters are tabulated in Table
\ref{tbl-2}, \ref{tbl-3}. We also measured the peak velocity ($V_{\mathrm{peak}}$), linewidth ($\triangle V$), and
main-beam temperature ($T_{\mathrm{mb}}$) of the six cores using the software GILDAS. The corresponding spectra of the
six cores are shown in Fig. \ref{core:spec}. The peak velocities of all cores but Core F coincide well with the systemic
velocity. However, the peak velocity of Core F is obviously blueshifted, which is attributed to the systemic variance
aforementioned.

\begin{deluxetable*}{lccccccccc}
\tabletypesize{\scriptsize} \tablecolumns{10} \tablewidth{0pc}
\tablecaption{Observed Parameters of $^{13}$CO Cores \label{tbl-2}}
\tablehead{
\colhead{} &\colhead{}&\colhead{}&\colhead{}    &  \multicolumn{3}{c}{$^{12}$CO (1-0)}    &
\multicolumn{3}{c}{$^{13}$CO (1-0)} \\
\cline{5-7} \cline{8-10} \\
\colhead{Name} & \colhead{R.A.\tablenotemark{a}}   & \colhead{DEC.\tablenotemark{a}}    & \colhead{$R_{eff}$\tablenotemark{a}} &\colhead{$V_{\mathrm{LSR}}$\tablenotemark{b}} &
\colhead{$\triangle V (^{12}\mathrm{CO})$\tablenotemark{b}}  & \colhead{$T_{\mathrm{mb}}(^{12}\mathrm{CO})$\tablenotemark{b}}& \colhead{$V_{\mathrm{peak}}$\tablenotemark{b}} &
\colhead{$\triangle V (^{13}\mathrm{CO})$\tablenotemark{b}}  & \colhead{$T_{\mathrm{mb}}(^{13}\mathrm{CO})$\tablenotemark{b}} \\
\colhead{ } & \colhead{(h m s)}   & \colhead{(d m s)}   & \colhead{(arcmin)} &\colhead{(km s$^{-1}$)} & \colhead{(km s$^{-1}$)}  & \colhead{(K)}&
\colhead{(km s$^{-1}$)} & \colhead{(km s$^{-1}$)}  & \colhead{(K)}}
\startdata
A & 18:33:56.18 & -7:36:49.2 & 1.3 & 95.49 (0.17) & 5.93 (0.17) & 14.0 & 95.30 (0.16) & 3.92 (0.16) & 6.9 \\
B & 18:33:54.26 & -7:37:32.4 & 0.9 & 94.67 (0.17) & 7.85 (0.17) & 12.0 & 94.21 (0.16) & 4.94 (0.16) & 6.3 \\
C & 18:33:47.06 & -7:39:42.0 & 1.1 & 94.78 (0.17) & 8.99 (0.17) & 12.0 & 94.61 (0.16) & 6.21 (0.18) & 4.0 \\
D & 18:33:39.86 & -7:39:49.2 & 1.3 & 94.73 (0.17) & 5.51 (0.17) & 13.7 & 94.45 (0.16) & 3.03 (0.16) & 7.2 \\
E & 18:33:34.58 & -7:39:06.0 & 1.4 & 94.83 (0.17) & 4.17 (0.17) & 18.9 & 94.63 (0.16) & 2.69 (0.16) & 9.7 \\
F & 18:33:25.46 & -7:36:27.6 & 0.6 & 97.07 (0.17) & 4.94 (0.17) & 10.1 & 97.18 (0.16) & 3.21 (0.21) & 3.4 \\
\enddata
\tablenotetext{a}{Parameters resulting from  the Clumpfind algorithm; Of these, $R_{eff}=\sqrt{(4A/\pi-\theta_{b}^{2})}/2$ is the deconvolved effective radius, where A is the projected area of
each core and $\theta_{b}$ is the beam width.}
\tablenotetext{b}{Parameters derived from the Gauss fitting in GILDAS.}
\end{deluxetable*}

Assuming conditions of local thermodynamic equilibrium (LTE) for molecular cores,
we derived the relationships of exciting temperature and H$_{2}$ column density as below:
\begin{equation}\label{eq2}
\begin{array}{rrr}
T_{\mathrm{r}}=\frac{\mathrm{h} \nu}{\mathrm{k}}[\frac{1}{J_{\nu}(T_{\mathrm{ex}})}-\frac{1}{J_{\nu}(T_{\mathrm{bg}})}]
$$[1-\mathrm{exp}(-\tau)]f\\
\end{array}
\end{equation}
\begin{equation}\label{eq3}
\begin{array}{rrr}
N_{^{13}\mathrm{CO}}=\frac{3\mathrm{k}}{8{\pi}^{3}B \mu^{2}}\frac{\mathrm{exp[hB}J(J+1)/\mathrm{k}T_{\mathrm{ex}}]}{(J+1)}
\times\frac{(T_{\mathrm{ex}}+\mathrm{hB}/3\mathrm{k})}{[1-\mathrm{exp(-h} \nu /k_{\mathrm{ex}})]}\int \tau_{\mathrm{13}}\mathrm{d}v
\end{array}
\end{equation}
where $J_{\nu}$ is defined as $\frac{1}{\mathrm{exp}(\mathrm{h} \nu/\mathrm{k}T)-1}$; $T_{r}$ is the brightness temperature
in units of K, $T_{\mathrm{ex}}$ is the exciting temperature, and $T_{\mathrm{bg}}=2.73$ K is the cosmic background radiation;
$\tau_{12}$ and $\tau_{13}$ are the optical depths of \co\ and \13co\ respectively;  $f$ is the fraction of the telescope
beam filled by emission, assumed to be 1; B and $\mu$ are the rotational constant and the permanent dipole moment of molecules
respectively; and $J$ is the rotational quantum number of the lower state in the observed transition. Given optical thick emission
for \co, the brightness temperatures of \co\ could yeild the corresponding exciting temperatures by the Equation 1. Substituting
$T_{ex}$ with Equation 1 and assuming optical thin emission for $^{13}$CO (1-0), Equation 2 becomes:
\begin{equation}\label{eq4}
\begin{array}{rrr}
N_{^{13}\mathrm{CO}}=\frac{3\mathrm{k}^{2}}{16{\pi}^{3} \mu^{2}\mathrm{h}\mathrm{B}^{2}}\frac{(T_{\mathrm{ex}}+\mathrm{hB}/\mathrm{3k})}{(J+1)^{2}}\mathrm{exp}[\mathrm{hB}J(J+1)/\mathrm{k}T_{\mathrm{ex}}]&\\
$$ \times\frac{1}{[1-\mathrm{exp}(-\mathrm{h} \nu /\mathrm{k}T_{\mathrm{ex}})][J_{\nu}(T_{\mathrm{ex}})-J_{\nu}(T_{\mathrm{bg}})]}\int T_{\mathrm{r}}(^{13}\mathrm{CO})\mathrm{d}v
\end{array}
\end{equation}
Given abundance ratios of $[^{12}\mathrm{CO}]/[\mathrm{H}_{2}]=8\times10^{-5}$ and $[^{12}\mathrm{CO}]/[^{13}\mathrm{CO}]=\tau_{12}/\tau_{13}=45$ \citep{sim01},
the column density $\mathrm{H}_{2}$  is $N(\mathrm{H}_{2})=5.6 \times 10^{5} N(^{13}\mathrm{CO})$.
If the core's structure is spherical, the mean $\mathrm{H}_{2}$ number density is $n(\mathrm{H}_{2})=N(\mathrm{H}_{2})/2R$,
and the core mass can be derived as:
\begin{equation}\label{eq5}
M_{\mathrm{core}}=\frac{4}{3}{\pi}R^{3}n(\mathrm{H}_{2})m_{\mathrm{H}}\mu
\end{equation}
where $m_{\mathrm{H}}$ is the mass of a hydrogen atom, and $\mu=2.8$ (e.g., \citet{sad13,kau08}) is
the mean atomic weight of gas. Table \ref{tbl-3} gives a summary of these derived parameters with errors
mainly dependent on the distance uncertainty of \g24. The mean column density, number density and mass of
the six cores are  $1.7\times10^{22}$ cm$^{-2}$, $1.6\times10^{3}$ cm$^{-3}$, $3.1 \times 10^{3}$ \msun,
respectively, while the estimated total mass of the shell could be $\sim2\times10^{4}$ \msun.
\begin{deluxetable*}{ccccccccccccccc}
\tabletypesize{\scriptsize} \tablecolumns{9} \tablewidth{0pc}
\tablecaption{Derived Parameters of $^{13}$CO Cores \label{tbl-3}}
\tablehead{\colhead{Name} & \colhead{R} & \colhead{$\tau_{12}$}& \colhead{$\tau_{13}$} &\colhead{$T_{\mathrm{ex}}$} &\colhead{$\int{T_{\mathrm{mb}}\mathrm{d}v}$\tablenotemark{a}}&
\colhead{$N_{^{13}\mathrm{CO}}$}& \colhead{$N_{\mathrm{H}_2}$}    & \colhead{$n_{\mathrm{H}_2}$} &
\colhead{$M_{\mathrm{LTE}}$}  \\
\colhead{} &\colhead{(pc)} & \colhead{}& \colhead{}& \colhead{(K)} &\colhead{(K km s$^{-1}$)}  & \colhead{(10$^{16}$ cm$^{-2}$)}  & \colhead{(10$^{22}$
cm$^{-2}$)} & \colhead{(10$^{3}$ cm$^{-3}$)} & \colhead{(10$^{3}M_{\odot}$)}  }

\startdata
A & 2.1 (0.4) & 30.4 & 0.7 & 16.6 & 31.6 (0.7) & 3.7(0.1) & 2.1 (0.1) & 1.6(0.3) & 4.4 (0.2)  \\
B & 1.5 (0.3) & 33.2 & 0.7 & 14.6 & 32.4 (0.7) & 3.6(0.1) & 2.0 (0.1) & 2.2(0.4) & 2.0 (0.7)  \\
C & 1.8 (0.3) & 17.9 & 0.4 & 14.6 & 26.6 (0.6) & 2.9(0.1) & 1.6 (0.1) & 1.5(0.3) & 2.5 (0.1)  \\
D & 2.1 (0.4) & 33.4 & 0.7 & 16.3 & 25.9 (0.5) & 3.0(0.1) & 1.7 (0.1) & 1.3(0.2) & 3.6 (0.1)  \\
E & 2.3 (0.4) & 32.6 & 0.7 & 21.5 & 30.4 (0.4) & 4.2(0.1) & 2.3 (0.1) & 1.7(0.3) & 5.7 (0.2)  \\
F & 1.0 (0.2) & 18.1 & 0.4 & 12.7 & 12.7 (0.6) & 1.3(0.1) & 0.7 (0.1) & 1.2(0.2) & 0.3 (0.1)  \\
\enddata
\tablenotetext{a}{The peak intensities of \13co\ resulting from the Clumpfind algorithm.}
\tablenotetext{b}{Errors mainly dependent on the uncertainty of the kinetic distance.}
\end{deluxetable*}

\subsection{Identification of Young Stellar Objects} \label{syso}
\begin{figure*}
 \centering
 \includegraphics[width=0.9\textwidth]{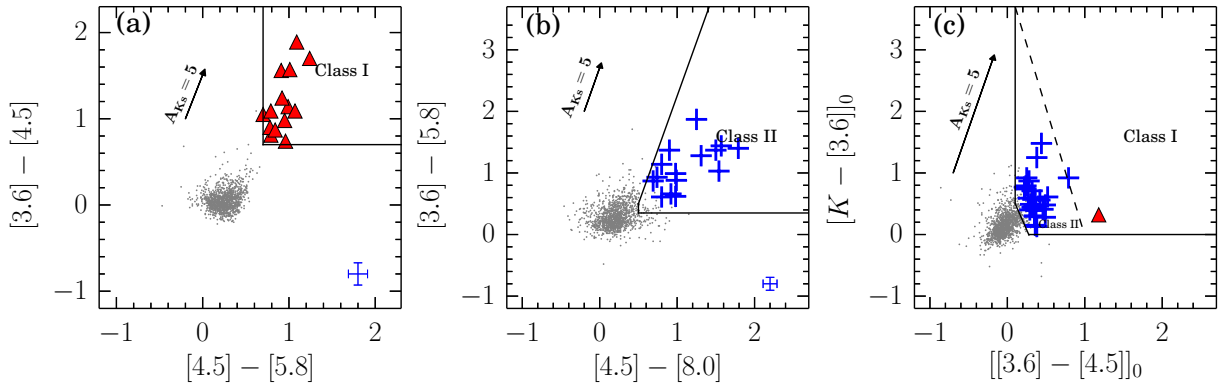}
\caption{Color-color diagrams used for identification of Class I and Class II YSOs. ``Red Triangle" represents
Class I YSOs and ``Blue Plus" denotes Class II YSOs, ``Filled gray circle" represents field stars.
The arrows show an extinction vector of $A_{K} = 5$ mag.}
\label{YSO}
\end{figure*}
\begin{figure}
  \centering
  \includegraphics[width=0.45\textwidth]{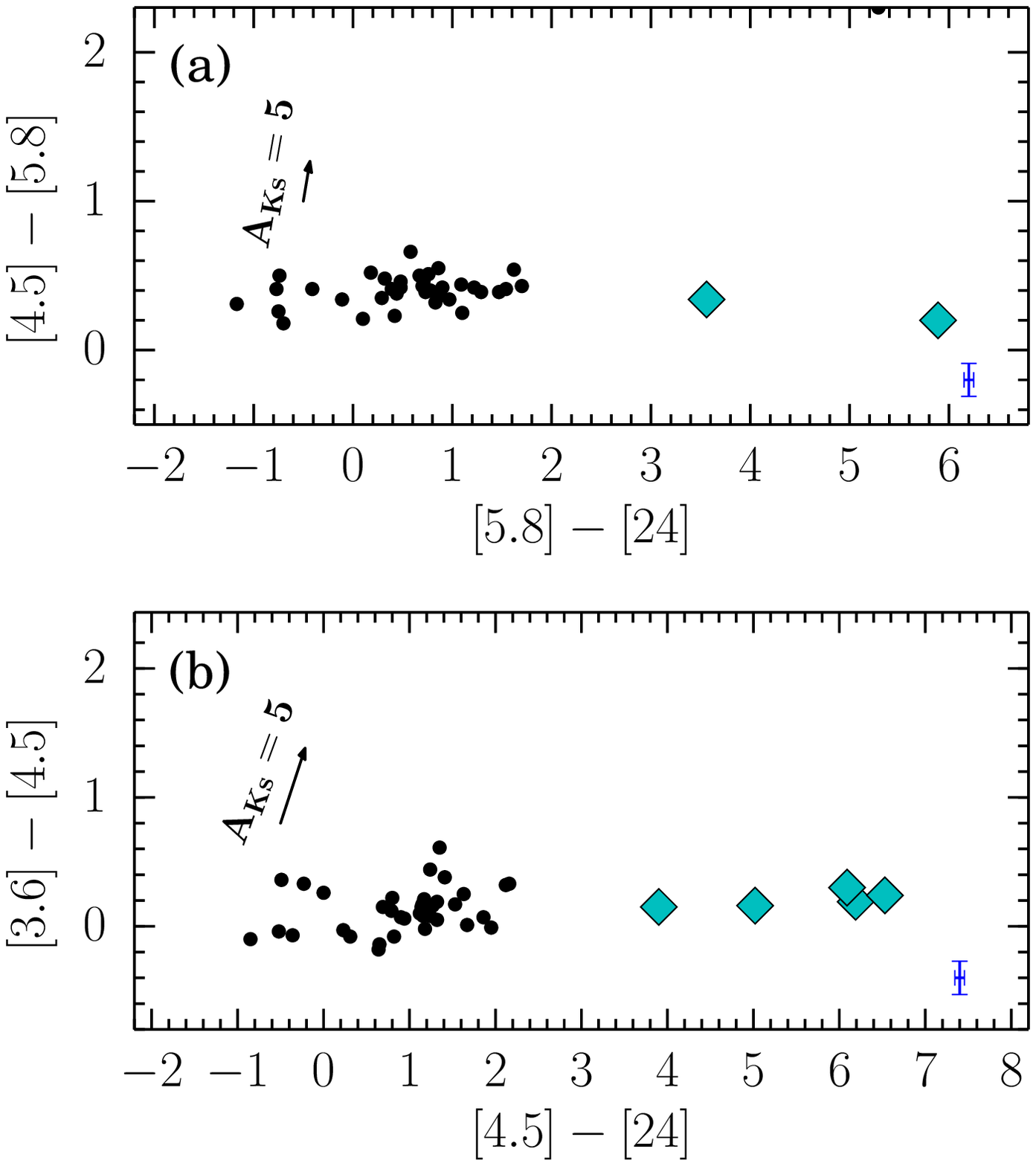}
\caption{Color-color diagrams used for identification of additional transition disks. ``Cyan Diamonds" represent the
transition disk YSOs, ``Filled black circles" represent field stars.
The arrows show an extinction vector of $A_{K} = 5$ mag.}
\label{mipsyso}
\end{figure}
 Infrared colors as a powerful proxy for measuring the excessive emission have been demonstrated by many
authors (e.g., \citet{all04, gut08}). Using the color selection scheme of \citet{gut09}, we identify YSOs
associated with the bubble in the light of  \emph{Spitzer} and UKIDSS data. The feasibility of this scheme
has been demonstrated in many star-forming regions located at different distances, such as W5-east NGC7538 \citep{cha14}
and NGC6634 \citep{wil13}. The procedure of isolating YSOs were carried out as below:
\begin{enumerate}
 \item We picked out 1239 sources having photometric uncertainties $\sigma < 0.2$ mag detections in all four IRAC bands.
 To ensure a confident YSO sample,  2 star-forming galaxies, 1 broad-line AGN, 1 knot of shock emission and 331
 sources with PAH-contaminated  apertures were removed by the criteria in various color spaces \citep{gut09}.
 The remaining uncontaminated sources are considered to be Class I YSOs, if  their colors follow:
 (1) $[4.5]-[5.8]>0.7$, and (2) $[3.6]-[4.5]>0.7$; and Class II YSOs, if their colors follow:
 (1) $[4.5]-[8.0]-\sigma_{24}>0.5$, (2)  $[3.6]-[5.8]-\sigma_{13}>0.35$,
 (3) $[3.6]-[5.8]+\sigma_{13}\leq \frac{0.14}{0.04} \times (([4.5]-[8.0]-\sigma_{24})-0.5)+0.5$,
 and (4) $[3.6]-[4.5]-\sigma_{12}>0.35$, where $\sigma_{24}, \sigma_{13}, \sigma_{12}$ are the combined
 measurement errors of two corresponding  IRAC bands (i.e., [4.5] and [5.8], [3.6] and [4.5], [3.6] and [4.5],
 respectively). In the end, 14 Class I and 15 Class II YSOs are obtained  by above constraints.
 These constraints are also shown in Fig. \ref{YSO}a and Fig. \ref{YSO}b.
 \item There are 8281 sources lacking detections at either 5.8 or 8.0 \um. We selected 2005 sources with
 $\sigma < 0.2$ mag detections at both 3.6 \um\ and 4.5 \um\ and with $\sigma < 0.1$ mag detections at
 least in UKIDSS near-infrared $H$ and $Ks$ bands. As suggested by \citet{gut09}, the line of sight extinctions
 of sources with only $H$ and $Ks$ bands detections are parameterized by the $E_{H-K}$ color excess, and sources
 with all UKIDSS bands detections are parameterized with the $\frac{E_{J-H}}{E_{H-K}}$ color excess ratio.  We
 then computed the adopted intrinsic colors from the measured photometries, including $[J-H]_{0}$, $[H-K]_{0}$,
 $[K-[3.6]]_{0}$ and $[[3.6]-[4.5]]_{0}$. Based on these dereddended colors, we identified additional YSOs
 candidates following  constraints: (1) $[[3.6]-[4.5]]_{0} - \sigma_{12} > 0.101$ (2) $[K-[3.6]]_{0} - \sigma_{K1} > 0$,
 and (3)  $[K-[3.6]]_{0} - \sigma_{K1} > -2.85714 \times ([[3.6]-[4.5]]_{0} - \sigma_{12} - 0.101) + 0.5$
 where $\sigma_{K1}$ is the  combined measurement error of $K$ and $[3.6]$.
 After further investigation, we found sources that obey
 $[K-[3.6]]_{0} - \sigma_{K1} > -2.85714 \times ([[3.6]-[4.5]]_{0} - \sigma_{12} - 0.401) + 1.7$ are likely
 protostars and Class II YSOs,  with
 the difference that Class II sources follow $[3.6]_{0} <14.5$, while  all protostars obey $[3.6]_{0} <15$. With
 this method, we obtain 1 Class I and 27 Class II YSOs. The corresponding constraints are displayed in Fig. \ref{YSO}c.
\item Combing objects with MIPS 24 \um\ detections with $\sigma<0.2$ mag, we isolated another 5 YSOs.  They are
transition disk sources, and classified as sources in an evolutionary stage between Class II and Class III \citep{gut09}.
They obey all the following criteria: (1) $[24]<7$, (2)  $[5.8]-[24]>2.5$ or $[4.5]-[24]>2.5$, (3) $[3.6]<14$.
Figure \ref{mipsyso} displays the corresponding criteria applied to the data.
\end{enumerate}

 As the result of the schemes described above, 62 YSO candidates were obtained. Combing them with 8 and 24 \um\
images, we found two compact sources not  identified as YSO candidates but are bright in the both 8 and 24 \um\ image.
One possible reason is that they are devoid of point-source photometries at longer wavelengths in IRAC bands due to
the diffuse emission. One of them, G024.1754+00.3994, is confirmed to be a compact HII region (see Sect. \ref{sCHIIregion}).
The other one, residing in the north of \g24, is considered to be a Class I protostar due to its bright emission at 24 \um.

 We note that some true YSOs might not be isolated since they reside in bright backgrounds (i.e., bright PAH emission),
 suggesting that the present number of YSO candidates may be a lower limit. The SED fitting to all point sources
 with detections in at least three wavelength bands (Liu et al., 2014, in preparation) would select more YSO candidates.
 Table \ref{tbl-5} gives a summary of 63 YSO candidates (17 Class I, 42 Class II and 5 transition disk). Figure \ref{inten}
 displays their spatial distributions overlaid on the \co\ and \13co\ velocity-integrated intensities. The majority of the YSO
 candidates projected on the region with moderate molecular gas indicates active star formation surrounding \g24.

\begin{figure}
  \centering
  \includegraphics[width=0.45\textwidth]{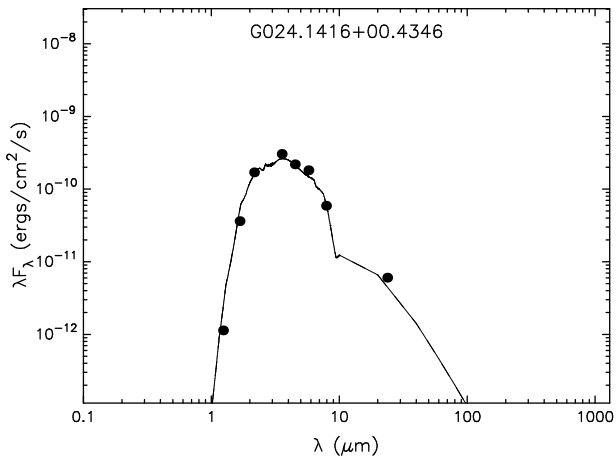}
\caption{Result of the fit to the candidate of exciting star using the on-line SED
fitting tool of \citet{rob06}. The filled circles show the input fluxes, and the black
line shows the best fit stellar photosphere model.}
\label{SED-exci}
\end{figure}

\section{Discussion}\label{sdiscu}
\begin{deluxetable*}{lccccccccccccc}
\tabletypesize{\scriptsize} \tablecolumns{11} \tablewidth{0pc}
\tablecaption{Observed Parameters of IRAS Sources \label{tbl-4}}
\tablehead{
\colhead{Name} & \colhead{R.A.}   & \colhead{DEC.}  & \colhead{$err_{rmaj}^{a}$}   & \colhead{$err_{rmin}^{a}$} & \colhead{$err_{PA}^{a}$} & \colhead{$F_{12}$} &\colhead{$F_{25}$} &\colhead{$F_{60}$} &\colhead{$F_{100}$}&
\colhead{$f_{qual}^{b}$} & \colhead{$log(\frac{F_{25}}{F_{12}})$}& \colhead{$log(\frac{F_{60}}{F_{12}})$}& \colhead{$L_{IR}^{c}$}\\
\colhead{ } & \colhead{(h m s)}   & \colhead{(d m s)} & \colhead{(arcsec)}   & \colhead{(arcsec)} & \colhead{(deg)} & \colhead{(Jy)} &\colhead{(Jy)} &\colhead{(Jy)} & \colhead{(Jy)}&
\colhead{}&\colhead{} &\colhead{}& \colhead{($10^{4}L_{\odot}$)}}
\startdata
18308-0741&18:33:35.859&-7:38:51.35 & 8 & 22 &86 &4.145  &7.804  &166.6 &560.2 &3232  &0.27  &1.60   &1.8  \\
18312-0738&18:33:57.318&-7:36:11.75 & 6 & 56 &85 &1.802  &4.806  &72.56 &396.1 &2332  &0.43   &1.60   &1.0  \\
\enddata
\tablenotetext{a}{The $95\%$ confidence error ellipse semimajor axis and position angle}
\tablenotetext{b}{The flux density quality: 2 represents moderate quality and 3 high quality.}
\tablenotetext{c}{The infrared luminosity is estimated (Casoli et al. (1986)) as follows: $F(10^{-13} W m^{-2})=1.75\times \left(\frac{F_{12}}{0.79}+\frac{F_{25}}{2}+\frac{F_{60}}{3.9}+\frac{F_{100}}{9.9}\right)$ and $L_{IR}=4\pi d^{2} F$.}
\end{deluxetable*}

\subsection{Star Formation Scenarios in Cores}\label{sstar:form:scena}
\begin{figure}
\centering
\includegraphics[width=0.48\textwidth]{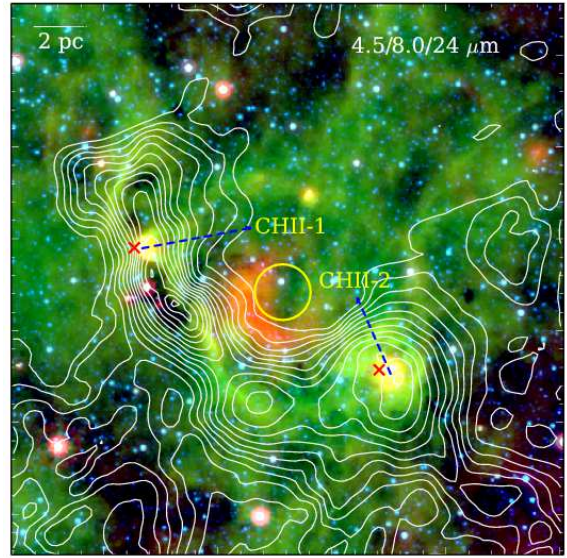}
\caption{Three-color composite image in 4.5 \um\ (blue), 8.0 \um\ (green) and 24 \um\ (red).
Two compact HII regions are labeled as CHII-1 and CHII-2.
The yellow circle represents a possible zone of existing ionizing sources.
The white contours and other symbols are the same as those in Fig. \ref{inten}.
In the center of the bubble, the 24 \um\ cavity is considered to be hot dust
broken by stellar winds. A scale bar of 2 pc is shown on the top left.}
\label{MIR}
\end{figure}

 The consistence of YSOs' spatial distribution with CO emission suggests that recently formed stars form within the molecular cloud and that star formation around the bubble is active. Additionally, the good match of dense molecular gas with PAH emission from PDRs and 24 \um\ emission from hot dust regions demonstrates the strong influence of IF created by the HII region on its surroundings.  Star formation around the bubble has to be affected by the expanding HII region. In what follows, we will discuss the star formation scenarios associated dense cores, coupled with their physical properties.

  \textbf{Core A} is located on the northeast of the bubble.
Within this core, $\sim8$ embedded YSO candidates (i.e., 1 Class I, 5 Class II, 1 transition disk, and CHII-1)
show IR excess emission in MIR or NIR wavelengths.
In Fig. \ref{MIR}, If the angular resolution was high enough, Core A would be divided
into a part coincided with an IR filament structure along the NW-SE direction and the
other part enclosing CHII-1 in the edge of the bubble.
Along the major axis of Core A (i.e., the direction NE-SW), emission of molecular gas of the core
shows the steep increase of intensities in the outskirts. In Fig. \ref{MIR}, some evolved stars
and YSOs possibly associated with this core are located in the front of it. Therefore, the large gradient of intensity
of \13co\ may be partially as a result of the external compression from them. It is plausible to partially attribute the broad line-width of
\13co\ in Core A, $3.92\pm0.16$ \vel, to the external pressure. Another star formation activities are also responsible for the broad line-width.
One IRAS source, 18312-0738, offsetting $45 \arcsec$ from the
peak of both Core A and CHII-1, is associated with the core. We note that this offsetting may result from a poor IRAS positoin
accuracy of $30\arcsec$. Following \citet{cas86}, the observed fluxes at 12, 25, 60 and 100 \um\ indicate an infrared
luminosity $L_{\mathrm{IR}}$ of $\sim2\times10^{4}$ $L_{\odot}$ (see Table \ref{tbl-4}), demonstrating a massive stellar system.
This is in great agreement with the existence of CHII-1. And hence, activities of massive star formation do proceed in Core A,
causing strong non-thermal motions and leading to the core, to some extent, with the broad line-width of \13co.

  \textbf{Core B} is located on the east of the bubble. Five embedded YSO candidates (three Class I and two Class II) within the core are indicative of presence of
active star formation in Core B. They may disturb their surrounding environments,
broadening the measured line-width of spectra. Thus, the broad line width of Core B, $\sim 4.94$ \vel, could be
partly attributed to the non-thermal motions driven by powerful activities of star formation.
In addition, the high LTE mass (a few of $10^3$ \msun) and intermediate number density (a few of $10^3$ cm$^{-3}$) of Core B show
a probability of forming a low-mass cluster or even high-mass protostars, which could also be supported by the
existence of two bright 24 \um\ point sources embedded within the core.
Figure \ref{MIR} shows that the IR counterpart of Core B is seen in absorption in both 8 \um\ and 24 \um\
against the infrared background. This indicates that molecular gas is located in front of the IR emission,
which is well consistent with the fact that this IR counterpart is an infrared dark cloud (IRDC) \citep{per09}.
The association of IRDC with Core B along with no detectable enclosed HII regions indicates that it is younger than Core A.

  \textbf{Core C} is located on the south periphery of the bubble. Around six YSO candidates (two Class I and four Class II) are associated with the core. This
shows the history of active ongoing star formation. The strong activities of star formation coupled with the effects of the
HII region on the Core C could produce non-thermal motions, resulting in the present broad line-width of \13co,
$\sim6$ \vel. In Comparison with Core B, there are no YSOs close to the emission peak of Core C, therefore,
we suggest that Core C may be younger than Core B.

  \textbf{Core D} is located on the southwest of the bubble. One embedded Class II YSO candidate offsets from the peak of the molecular core.
We note that some true YSOs may be missed due to the limitation of the color-color scheme aforementioned in Sect. \ref{syso}.
The broad line width ($3.03\pm0.16$ \vel), high number density ($1.3\pm0.2 \times 10^{3}$ cm$^{-3}$) and mass ($3.6\pm0.1\times10^{3}$ \msun) of Core D are
comparable to those of three cores aforementioned. Given Core D not associated with either any bright 24 \um\ point sources or strong PDRs,
it would be at an early stage or in the process or in the process of low-mass star formation.

  \textbf{Core E} is located on the southwest of the bubble. About seven YSO (four Class I and three Class II) candidates are associated with the core.
Six of them distribute sparsely in the outer edge of the core. This may be inconsistent with the actual situations due to the
missing problem of identifying YSOs mentioned above. In the center of Core E, one Class I object and one bright IRAS source
(i.e., 18308-0741) are detected. The IRAS source has an estimated luminosity of $\sim10^{4}$ $L_{\odot}$. This is a good indicator
of a massive stellar system, in a good agreement with the fact that CHII-2 has formed beginning
to ionize the neighbor ISMs and sculpt a small photoionized bubble of gas embedded within the cold, dense, and massive molecular core (see Sect. \ref{sCHIIregion}).
With energetic activities of massive star formation developing, inevitable non-thermal
disturbances have to be emerged. Therefore, the broad linewidth of Core E, $\sim2.69$ \vel, could be partially interpreted as broadening
of the disturbances. Due to the existence of the compact ionized region in Core A and Core E, their evolved lifetimes are comparable.

  \textbf{Core F} is located on the west of the bubble. A total of $\sim3$ YSO candidates (two Class II and one transition disk) distribute along the edge of the core,
witnessing the star formation history. It is the diffuse molecular entity with the lowest column density, number
density and LTE mass among all cores. Given the core lying away from the bubble (see Fig. \ref{MIR}), effects of the
classical HII region on Core F may be weaker than on other cores.

\subsection{Exciting Stars} \label{Sexcitingstar}
To investigate the main exciting star, we calculate the number of UV ionizing photons per second ($N_{\mathrm{uv}}$)
from the free-free emission at 20 cm, based on the following equation \citep{cha76}:
\begin{equation}\label{eq1}
N_{\mathrm{uv}}=0.76 \times 10^{47} T_{4}^{-0.45}\nu_{\mathrm{GHz}}^{0.1}S_{\nu}d_{\mathrm{kpc}}^{2}
\end{equation}
where $T_{4}$ is the electron temperature in units of $10^{4}$ K, $S_{\nu}$ is the measured flux density in units
of Jy, $\nu_{\mathrm{GHz}}$ is the frequency in units of GHz and $d_{\mathrm{kpc}}$ is the distance in units of kpc.
If the HII region reaches the equilibrium at an electron temperature of $T = 10^{4}$ K and a total flux density
of $0.46\pm0.05$ Jy at 20 cm (measured by integrating over the $4\sigma$ ($\sigma=$0.15 mJy/beam) contour within
the radius of the HII region), the estimated value of $N_{\mathrm{uv}}$ is $1.15\pm0.04 \times 10^{48}$ ph s$^{-1}$.
In this calculation, any ionizing photons absorbed by dust or running away from the bubble were not accounted for,
and hence, the present $N_{\mathrm{uv}}$ has been underestimated. Similar argument is made by \citet{wat08}, on the
basis of studies of several bubbles, that $N_{\mathrm{uv}}$ estimated with this method is statistically lower than
the expected value by about a factor of 2. In the following, we adopt $\sim2.3 \times 10^{48}$ ph s$^{-1}$ as a
reference value. It corresponds to a spectral type of  $\sim$O8V  (see Table 4 of \citet{mar05}). Given uncertainties
aforementioned, the estimated spectral type should be considered with a caveat.

Following the method of searching for exciting stars \citep{pom09}, sources with all detections in NIR and IRAC bands
were extracted from a region centered at the hot dust cavity seen at 24 \um\ (i.e., the yellow solid circle in Fig. \ref{MIR}).
The SED fittings to these sources give one potential candidate, G024.1416+00.4364. Its result of SED fitting is displayed in
Fig. \ref{SED-exci}, deriving a temperature of $\sim3.5 \times 10^{4}$ K and a mass of $\sim 20$ \msun. These values are
consistent with stellar parameters of $\sim$O8V \citep{mar05}. Therefore, G0241416+00.4364 would be a major exciting
star candidate which could provide enough energy to create the dust bubble and regulate its surroundings.

\subsection{Two Compact HII regions} \label{sCHIIregion}
\begin{figure}
\centering
\includegraphics[width=0.48\textwidth]{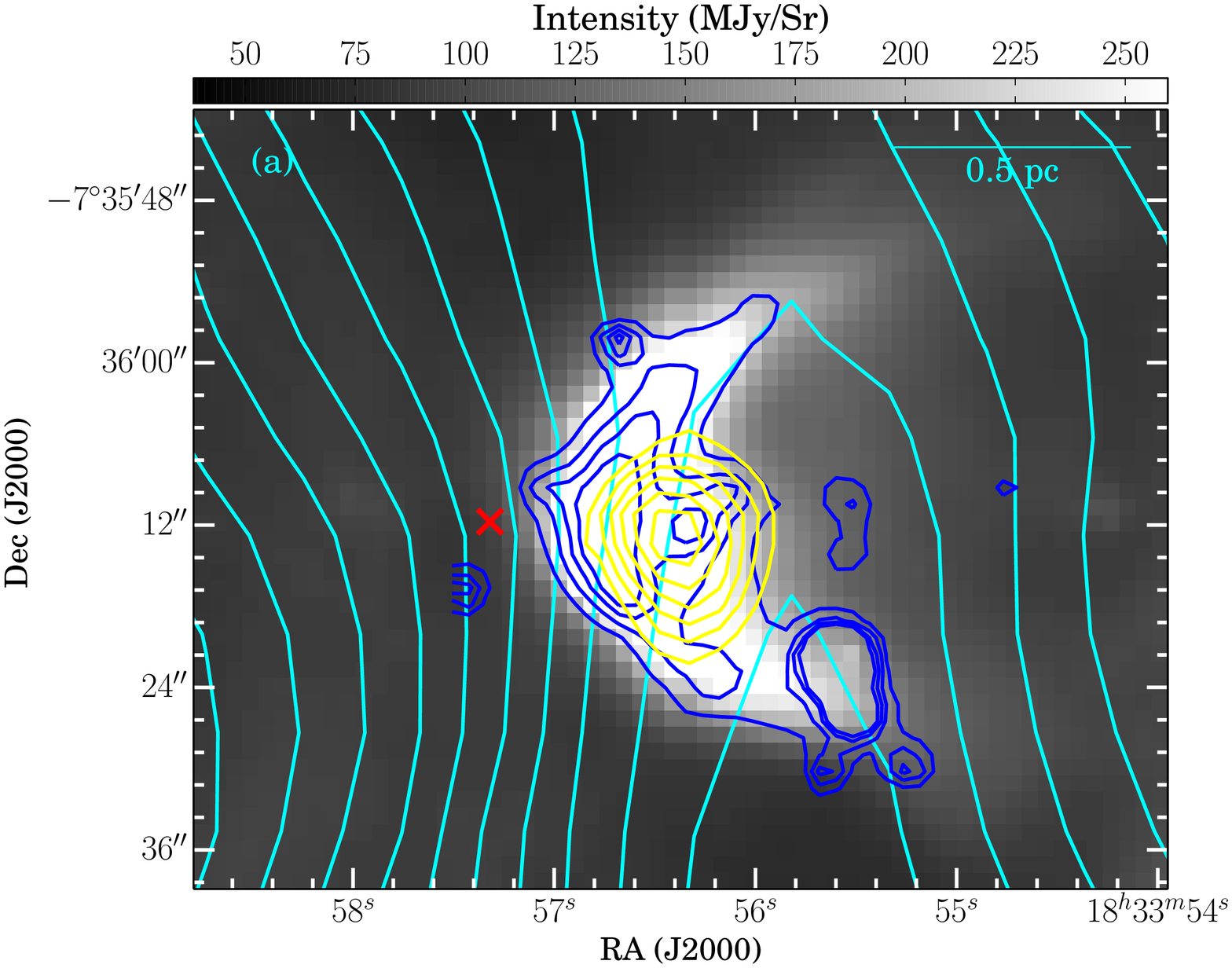}
\includegraphics[width=0.48\textwidth]{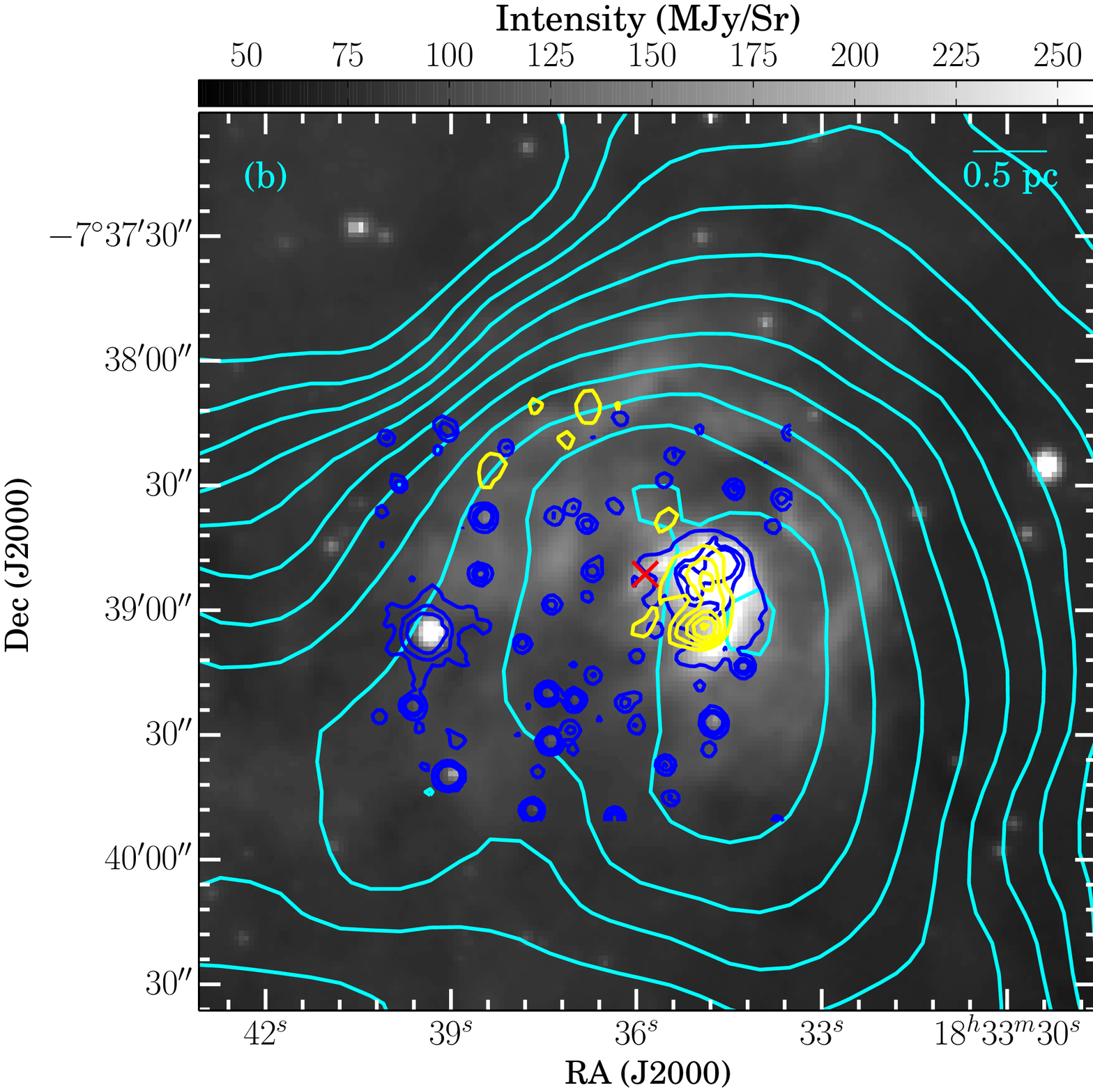}
\caption{(a): 8.0 \um\ image of the compact HII region, CHII-1, overlaid with contours of the radio continuum emission at 20 cm in yellow, 4.5 \um\ emission in blue,
and \13co\ emission in cyan. The contours start from 0.9 mJy/pixel with a step of 0.2 mJy/pixel for the 20 cm emission,
from 8 MJy/sr with a step of 4 MJy/sr for the 4.5 \um\ emission and from 19.6 K \vel\ with a step of 0.7 K \vel\ for the \13co\ intensity. The red cross represents
the IRAS source 18308-0741. A scale bar of 0.5 pc is shown on the top right. (b): Same for the compact HII region, CHII-2. The contours start from 0.8 mJy/pixel with a step of 0.2 mJy/pixel for the 20 cm emission,
from 8 MJy/sr to 16 MJy/sr and finally to 28 MJy/sr for the 4.5 \um\ emission and from 7 K \vel\ with a step of 0.7 K \vel\  for the \13co\ intensity. The red cross represents
the IRAS source 18312-0738. A scale bar of 0.5 pc is shown on the top right.}
\label{UCHII}
\end{figure}

 Figure \ref{MIR} displays the two compact sources, G024.1754+00.3994
and G024.0953+00.4579 denoted as CHII-1 and CHII-2 respectively. These two sources
are clearly associated with thermal free-free continuum at 20 cm, PAH emission from the PDRs, and hot dust emission out of thermal
equilibrium seen at 24 \um. In existing studies (e.g., \citet{urq13}, and reference therein), these observable features have always been
adopted as useful criteria to identify ultracompact (UC) or compact HII regions. They are generally of different
scale sizes and particle densities: $D\leq0.1$ pc and $n\geq10^{4}$ cm$^{-3}$ for UCHII regions;
$D\leq0.5$ pc and $n\geq10^{3}$ cm$^{-3}$ for CHII regions \citep{kur05}. The free-free emission confines
the size of CHII-1 and CHII-2 to be 0.4 pc and 0.8 pc
respectively. In combination of their number density of $\sim10^{3}$ cm$^{-3}$ (assumed to be similar to those of their hosting
molecular cores), they could be plausibly identified as CHII candidates. These two CHII candidates coupled with the classic HII region inside the bubble
exhibit a hierarchical structure, indicative of secondary star formation on the rim of the bubble.

 CHII-1 and CHII-2 have flux densities of $2.7\pm0.3$ mJy and $8.0\pm0.2$ mJy at 20 cm respectively.
The minimum ionizing photon flux required to maintain ionization of the region is log $N_{Lyc} > 45.8$  for CHII-1 and  $> 46.3$
for CHII-2, both corresponding to a B0.5V star \citep{pan73}. In the two CHII sources, the 4.5 \um\ emission
tracing shocked molecular gas \citep{cyg08} is spatially coincided with the bright PAH emission and consistent with the free-free
emission. This sheds light on the strong influence of ionizing sources on their natal molecular cloud (i.e., Core A and Core E), indicating the possibility of
existing outflows from massive stars. Of greater interest is the CHII-1, as its morphology appears as a cometary object seen in
both the 4.5 and 8.0 \um\ image. High resolution observations of these two sources would help to determine their nature
and origin.

\subsection{Feedback From Massive Stars}\label{trigger}
 \g24\ is an almost perfectly spherical bubble around the HII region ionized by an $\sim$O8V star(s). The PDRs just
outside the IF suggests that the strong influence of the HII region on its adjacent molecular cloud. This influence could be
partially supported by shocked compressed material at the outskirts of the shell-like molecular cloud. In \g24\ the hot
dust cavity broken along the northwest indicates that the stellar winds from an late-O star (stars) probably play a critical
role in affecting its associated cloud. The feedback of stellar winds has also been proposed in the study of the bubbles N10,
N21, N49 \citep{wat08}.

 On the one hand, \citet{deh12} suggested that during the evolution of an HII region its IF reaches regions of lower density than those
of higher density and the ionization occurs more quickly in the former case. In diffuse region of lower densities, no neutral material might be
collected, which applies well to \g24. The diffuse and rarified molecular gas associated with diffuse PAH emission to the
northwest of \g24\ indicates the original mediums of low densities and the photoionization of the HII region. Additionally,
few young stars means that star formation to the northwest of \g24\ might be halted due to disruptions of molecular gas by
the expansion of the HII region and stellar winds. On the other hand, neutral material seems to be collected into the
border on the southeast of \g24. The consistence of YSOs' spatial distribution with the collected layer shows active
star formation on the border of the bubble. Additionally, the hierarchies of secondary star formation on the rim of the bubble
(see Sect. \ref{sCHIIregion}) indicates a promising sign for triggering star formation \citep{oey05,zav06,deh10,elm11,sim12,sam14}.

\subsection{Collect and Collapse Process}
 The morphology of the cloud projected on the periphery of \g24\ indicates that the shocked molecular layer is presumably swept up and
assembled  during the bubble's expansion, whereby we discuss star formation within the bubble in the context of the collect and collapse
process.

The dynamical age of the HII region can be simply estimated as follows \citep{dys80}:
\begin{eqnarray}
\nonumber
R_{s} &=& (3N_{\mathrm{uv}}/4\pi n_{0}^{2}\alpha_{B})^{1/3} \\
 t_{\mathrm{dyn}}&=&(\frac{4R_{s}}{7c_{s}})[(\frac{R}{R_{s}})^{7/4}-1]
\end{eqnarray}
 where $R_{s}$ is the radius of the Stromgren sphere in units of cm, $n_{0}$ is the initial particle density of ambient neutral gas in
units of cm$^{-3}$, $\alpha_{B}$ = $2.6\times10^{-13}(10^{4} \mathrm{K}/T)^{0.7}$ cm$^{3}$ s$^{-1}$ \citep{kwa97} is the the coefficient
of the radiative recombination, c$_{s}$ is the isothermal sound speed of ionized gas assumed to be
10 km s$^{-1}$, $R$ is the radius of the HII region in units of cm. To obtain an approximate $n_{0}$, we assume that the total mass of the
molecular shell (i.e., $2\times10^{4}$ \msun) was distributed homogeneously in a sphere of radius 3.9 pc (i.e., the bubble radius), which gives an estimated
density of $\sim1.3\times10^{3}$ cm$^{-3}$. This value may be just a lower limit since this estimate does not consider the mass of the ionized gas within the bubble.
Therefore, we adopt the average density over the six cores (i.e., $\sim1.6\times10^{3}$ cm$^{-3}$) as a possible upper limit.
If \g24\ formed and evolved in a uniform density of $(1.3-1.6)\times10^{3}$ cm$^{-3}$,
the bubble's radius of $3.7\pm0.7$ pc derives dynamical ages of $1.5-1.6$ Myr. We note that the
the realistic evolution of the HII region is not in a strictly uniform medium, therefore, the estimated age is very uncertain,
and should be considered with a caution.

The analytical model of \citet{whi94} describes fragmentation time of the
shocked dense layer surrounding expanding HII regions. The time at which the fragmentation commences is calculated as follows:
\begin{equation}\label{eqa11}
    t_{\mathrm{frag}}=1.56 a_{.2}{}^{7/11} N _{49}{}^{-1/11} n_3{}^{-5/11}
\end{equation}
where $a_{0.2}$ is the sound speed inside the shocked layer in unites of 0.2 \vel, $N_{49}$ is the ionizing photon flux in units
of $10^{49}$ ph $s^{-1}$, and $n_{3}$ is the initial gas atomic number density in units of $10^{3}$ cm$^{-3}$.
The above calculation shows $t_{frag}$ depends weakly on $N _{uv}$,
somewhat on $n_{0}$ and greatly on $a_{s}$. We consider the thermal sound speed in the collected layer  to
be $\sim0.2$ \vel at an average temperature of $\sim16$ K over the shell, indicating fragmentation time of $\sim1.7$ Myr
for the density of $\sim1.3\times10^{3}$ cm$^{-3}$ and $\sim1.5$ Myr for the density of $\sim1.6\times10^{3}$ respectively.
 In the estimate of fragmentation time, the stellar
winds have not been took into account due to lacking their mechanical luminosity. Therefore, fragmentation time derived
from Eq. \ref{eqa11} is overestimated. Given the uncertainty, this time scale is comparable to or even
smaller than the dynamical age of the HII region. This demonstrates that during the lifetime of \g24 the collected
 molecular cloud has enough time to become gravitationally unstable to fragment into molecular cores.
 This process follows the model of ``collect and collapse", which is indicative of triggering.

 To conclude, a combination of the enhanced number of candidate YSOs as well as secondary star formation in the collected layer
 and time scales involved ($t_{\mathrm{dyn}},t_{\mathrm{frag}}$)
 indicates a possible scenario of triggered star formation towards \g24\ by both the expansion of the bubble and stellar winds,
 signified by the ``collect and collapse" process.

  While the conclusion on triggering in \g24\ is conservative, it remains to be further confirmed because of at least three difficulties.
 First, it is hard to discern which stars are triggered and which stars formed spontaneously \citep{elm11,dal11,dal12,dal13}. Second, the YSOs surface density enhancement
 on the rim of the bubble might be re-distributed by its expansion or forming in situ \citep{elm11}. Finally, the incompleteness
 of YSO candidates aforementioned in Sect. \ref{syso} hardly restore their true spatial distribution. All these problems hinder us
 from giving any convincing evidence on triggering. In later work of Liu et al.
 (2014, in preparation), we will run SED fitting to 10509 sources to select more completed YSO candidates. Analyzing their age differences
 (in an attempt to search for possible age gradients) would be useful to remit above difficulties.

\section{Summary}
We have performed an extensive study of the IR bubble \g24 using J=1-0 lines of CO and its
isotopologues, infrared and radio data. The main results are summarized as follows:

\begin{enumerate}

\item The molecular gas around the bubble has a velocity of $94.8 \pm 0.42$ \vel. Its emission is prominently in the
southeast of the bubble, sculpted as a shell with a mass of $\sim2\times10^{3}$ \msun. It is presumably assembled during
the expansion of the bubble.
\item The expanding shell consists of six dense cores. Their dense (a few of $10^{3}$ cm$^{-3}$) and massive (a few of $10^{3}$ \msun)
properties coupled with the broad linewidths ($>2.5$ \vel) suggest they are promising sites of forming high-mass stars or clusters.
This, in fact, could be further consolidated by the two compact HII regions embedded within Core A and Core E.
\item We tentatively identified and classified 63 YSO candidates (16 Class I, 42 Class II and 5 Transition-disk objects) by the color scheme.
They are predominantly projected on the regions with strong emission of molecular gas, showing active star formation especially in the shell.
\item \g24\ surrounds an almost spherical HII region. It is ionized by an $\sim$O8V star(s), of the dynamical age $\sim$1.6 Myr. The PDRs just outside
the IF and the dust cavity demonstrate the strong influence of both HII region and stellar winds on the adjacent environments and star formation.
\item From the enhanced number of YSO candidates and secondary star formation in the shell, we suggest that star formation in
the shell might be triggered by a combination of the expanding HII region and stellar winds. The comparison between the estimated dynamical age of the HII region and
the fragmentation time of the shell further indicates that triggered star formation might function through the ``collect and collapse" process.
\end{enumerate}

\begin{acknowledgements}
   We thank the anonymous referee for comments that much improve the quality of this paper.
This work is supported by the National Natural Science Foundation of China through grant NSFC 11073027, 11373009 and 11433008, the Ministry of Science
and Technology of China through grants 2012CB821800 (a State Key Development Program for Basic Research)
and 2010DFA02710 (by the Department of International Cooperation of MOST). We are grateful to the staffs at the Qinghai
Station of PMO for their hospitality and assistance during the observations. We thank the Key Laboratory for Radio Astronomy, CAS,
for partial support in the operation of the telescope.
\end{acknowledgements}



\clearpage
\LongTables
\begin{landscape}
\begin{deluxetable*}{llllllllllllll}
\centering
\tabletypesize{\scriptsize(4pt)} \setlength{\tabcolsep}{0.01in}
\tablecolumns{14}
\tablewidth{0pt}
\tablecaption{Results of YSO Classification. \label{tbl-5}}
\tablehead{\colhead{Number}&\colhead{Designation}&\colhead{R.A.}&\colhead{DEC.}&\colhead{$J$}&\colhead{$H$}& \colhead{$K$}&\colhead{$3.6$}
&\colhead{$4.5$}& \colhead{$5.8$}&\colhead{$8.0$}&\colhead{$24$}&\colhead{Type}&\colhead{Associations$^{\dag}$}\\
\colhead{ }&\colhead{ }&\colhead{(deg)}&\colhead{(deg)}&\colhead{(mag)}&\colhead{(mag)}& \colhead{(mag)}  & \colhead{(mag)} & \colhead{(mag)}
&\colhead{(mag)}&\colhead{(mag)}&\colhead{(mag)}&\colhead{ }&\colhead{ }}
\startdata

         1 &          G024.1244+00.5352 &  18h33m21.316s  & -7d35m07.46s   &  14.55    (   0.05)   &   12.18       (0.05)    &  10.97      ( 0.03)    &   9.88      ( 0.03)   &    9.68      ( 0.03)   &    9.26      ( 0.03)   &    8.70      ( 0.02)   &    6.99      ( 0.02)  &                 II     &       F     \\
         2 &          G024.0286+00.4726 &  18h33m24.076s  & -7d41m57.63s   &    nan    (    nan)   &     nan       ( nan)    &    nan      (  nan)    &  12.42      ( 0.06)   &   11.94      ( 0.09)   &   11.43      ( 0.09)   &   10.95      ( 0.11)   &     nan      (  nan)  &                 II     &              \\
         3 &          G024.0233+00.4617 &  18h33m25.823s  & -7d42m32.63s   &  15.60    (   0.09)   &   14.29       (0.08)    &  13.48      ( 0.06)    &  12.70      ( 0.09)   &   12.28      ( 0.09)   &     nan      (  nan)   &     nan      (  nan)   &     nan      (  nan)  &                 II     &              \\
         4 &          G024.0219+00.4606 &  18h33m25.902s  & -7d42m38.76s   &    nan    (    nan)   &     nan       ( nan)    &    nan      (  nan)    &  14.41      ( 0.14)   &   12.86      ( 0.10)   &   11.95      ( 0.14)   &     nan      (  nan)   &     nan      (  nan)  &                  I     &              \\
         5 &          G024.0612+00.4757 &  18h33m27.045s  & -7d40m08.33s   &    nan    (    nan)   &     nan       ( nan)    &  13.09      ( 0.05)    &  10.54      ( 0.04)   &    9.56      ( 0.04)   &    8.62      ( 0.03)   &    7.71      ( 0.02)   &    5.93      ( 0.02)  &                  I     &       E     \\
         6 &          G024.1028+00.4953 &  18h33m27.489s  & -7d37m22.84s   &    nan    (    nan)   &   14.41       (0.07)    &  13.73      ( 0.08)    &  13.10      ( 0.06)   &   12.75      ( 0.09)   &     nan      (  nan)   &     nan      (  nan)   &     nan      (  nan)  &                 II     &       E       \\
         7 &          G024.1361+00.5104 &  18h33m27.956s  & -7d35m11.34s   &    nan    (    nan)   &     nan       ( nan)    &    nan      (  nan)    &  13.27      ( 0.06)   &   13.09      ( 0.08)   &     nan      (  nan)   &     nan      (  nan)   &    6.89      ( 0.04)  &                 TD     &       F     \\
         8 &          G024.1102+00.4905 &  18h33m29.328s  & -7d37m07.17s   &    nan    (    nan)   &   14.49       (0.06)    &  13.21      ( 0.04)    &  12.27      ( 0.06)   &   11.74      ( 0.08)   &   11.74      ( 0.12)   &     nan      (  nan)   &     nan      (  nan)  &                 II     &       E       \\
         9 &          G024.0696+00.4656 &  18h33m30.166s  & -7d39m58.15s   &    nan    (    nan)   &     nan       ( nan)    &    nan      (  nan)    &  10.51      ( 0.03)   &    9.61      ( 0.03)   &    8.84      ( 0.03)   &    9.06      ( 0.03)   &     nan      (  nan)  &                  I     &       E     \\
        10 &          G024.1283+00.4927 &  18h33m30.869s  & -7d36m05.78s   &  13.66    (   0.03)   &   13.10       (0.03)    &  12.65      ( 0.03)    &  12.01      ( 0.07)   &   11.82      ( 0.07)   &   11.59      ( 0.11)   &     nan      (  nan)   &     nan      (  nan)  &                 II     &       F     \\
        11 &          G024.0531+00.4496 &  18h33m31.747s  & -7d41m17.57s   &  14.83    (   0.04)   &   14.30       (0.04)    &  13.99      ( 0.08)    &  13.77      ( 0.07)   &   12.61      ( 0.11)   &     nan      (  nan)   &     nan      (  nan)   &     nan      (  nan)  &                  I     &       E     \\
        12 &          G024.1231+00.4796 &  18h33m33.125s  & -7d36m43.97s   &  13.54    (   0.03)   &   10.08       (0.02)    &   8.20      ( 0.03)    &   6.71      ( 0.03)   &    6.32      ( 0.03)   &    5.78      ( 0.02)   &    5.58      ( 0.02)   &    4.92      ( 0.01)  &                 II     &       E       \\
        13 &          G024.1689+00.5021 &  18h33m33.396s  & -7d33m40.50s   &    nan    (    nan)   &     nan       ( nan)    &    nan      (  nan)    &  13.88      ( 0.10)   &   12.78      ( 0.08)   &   11.71      ( 0.10)   &   11.80      ( 0.14)   &    8.32      ( 0.03)  &                  I     &              \\
        14 &          G024.0953+00.4579 &  18h33m34.670s  & -7d38m49.04s   &  15.70    (   0.08)   &   12.72       (0.06)    &  10.67      ( 0.03)    &   8.54      ( 0.09)   &    7.81      ( 0.05)   &    6.84      ( 0.06)   &    6.08      ( 0.14)   &     nan      (  nan)  &                  I     &       E     \\
        15 &          G024.1554+00.4863 &  18h33m35.286s  & -7d34m49.72s   &  14.73    (   0.06)   &   13.90       (0.08)    &  13.29      ( 0.06)    &  12.85      ( 0.05)   &   12.52      ( 0.07)   &   12.31      ( 0.18)   &     nan      (  nan)   &     nan      (  nan)  &                 II     &              \\
        16 &          G024.0378+00.4081 &  18h33m38.959s  & -7d43m15.37s   &  14.78    (   0.04)   &   14.16       (0.05)    &  13.79      ( 0.07)    &  13.56      ( 0.09)   &   13.23      ( 0.08)   &     nan      (  nan)   &     nan      (  nan)   &     nan      (  nan)  &                 II     &              \\
        17 &          G024.1022+00.4304 &  18h33m41.361s  & -7d39m12.69s   &    nan    (    nan)   &     nan       ( nan)    &    nan      (  nan)    &  11.84      ( 0.04)   &   11.48      ( 0.06)   &   10.96      ( 0.07)   &   10.50      ( 0.18)   &     nan      (  nan)  &                 II     &       D     \\
        18 &          G024.1629+00.4584*&  18h33m42.121s  & -7d35m12.10s   &    nan    (    nan)   &    nan        ( nan)    &    nan      ( nan)     &  13.94      ( 0.09)   &   13.55      ( 0.16)   &     nan      (  nan)   &     nan      (  nan)   &     nan      (  nan)  &                 I      &             \\
        19 &          G024.2168+00.4853 &  18h33m42.363s  & -7d31m35.32s   &  14.18    (   0.04)   &   13.59       (0.05)    &  13.29      ( 0.05)    &  12.95      ( 0.06)   &   12.60      ( 0.12)   &     nan      (  nan)   &     nan      (  nan)   &     nan      (  nan)  &                 II     &              \\
        20 &          G024.0813+00.4141 &  18h33m42.514s  & -7d40m46.32s   &    nan    (    nan)   &     nan       ( nan)    &  13.61      ( 0.04)    &  12.18      ( 0.05)   &   11.91      ( 0.05)   &   11.51      ( 0.08)   &   10.99      ( 0.17)   &     nan      (  nan)  &                 II     &       C     \\
        21 &          G024.1080+00.4241 &  18h33m43.353s  & -7d39m04.48s   &  15.25    (   0.05)   &   14.47       (0.05)    &  13.64      ( 0.07)    &  12.29      ( 0.05)   &   11.89      ( 0.07)   &   11.77      ( 0.09)   &     nan      (  nan)   &     nan      (  nan)  &                 II     &       C     \\
        22 &          G024.1121+00.4182 &  18h33m45.088s  & -7d39m01.28s   &    nan    (    nan)   &     nan       ( nan)    &    nan      (  nan)    &  13.73      ( 0.07)   &   11.84      ( 0.07)   &   10.75      ( 0.06)   &   10.71      ( 0.15)   &    6.39      ( 0.02)  &                  I     &       C     \\
        23 &          G024.0695+00.3902 &  18h33m46.342s  & -7d42m03.70s   &  10.75    (   0.03)   &    7.99       (0.04)    &   6.35      ( 0.02)    &   5.36      ( 0.16)   &    4.79      ( 0.04)   &    4.22      ( 0.03)   &    3.99      ( 0.06)   &    1.69      ( 0.03)  &                 II     &              \\
        24 &          G024.1304+00.4180 &  18h33m47.169s  & -7d38m03.09s   &    nan    (    nan)   &     nan       ( nan)    &    nan      (  nan)    &  12.53      ( 0.08)   &   11.66      ( 0.06)   &   10.82      ( 0.07)   &   10.45      ( 0.06)   &     nan      (  nan)  &                  I     &       C       \\
        25 &          G024.1650+00.4325 &  18h33m47.911s  & -7d35m48.33s   &  15.02    (   0.06)   &   14.13       (0.07)    &  13.71      ( 0.07)    &  12.90      ( 0.07)   &   12.65      ( 0.09)   &     nan      (  nan)   &     nan      (  nan)   &     nan      (  nan)  &                 II     &              \\
        26 &          G024.1882+00.4411 &  18h33m48.660s  & -7d34m20.25s   &    nan    (    nan)   &   14.72       (0.06)    &  13.95      ( 0.06)    &  12.90      ( 0.06)   &   12.57      ( 0.08)   &   12.44      ( 0.17)   &     nan      (  nan)   &     nan      (  nan)  &                 II     &              \\
        27 &          G024.1192+00.4014 &  18h33m49.492s  & -7d39m06.45s   &  15.45    (   0.06)   &   13.99       (0.07)    &  13.29      ( 0.06)    &  12.85      ( 0.06)   &   12.38      ( 0.08)   &     nan      (  nan)   &     nan      (  nan)   &     nan      (  nan)  &                 II     &       C     \\
        28 &          G024.2132+00.4498 &  18h33m49.584s  & -7d32m45.81s   &    nan    (    nan)   &   12.48       (0.03)    &   9.98      ( 0.02)    &   6.83      ( 0.04)   &    5.75      ( 0.04)   &    4.96      ( 0.02)   &    4.30      ( 0.02)   &    2.44      ( 0.03)  &                  I     &              \\
        29 &          G024.0762+00.3783 &  18h33m49.650s  & -7d42m02.08s   &  15.16    (   0.05)   &   14.36       (0.06)    &  13.89      ( 0.07)    &  13.48      ( 0.11)   &   13.01      ( 0.12)   &     nan      (  nan)   &     nan      (  nan)   &     nan      (  nan)  &                 II     &              \\
        30 &          G024.1183+00.3868 &  18h33m52.529s  & -7d39m33.46s   &    nan    (    nan)   &   12.89       (0.03)    &  10.12      ( 0.03)    &   7.91      ( 0.03)   &    7.67      ( 0.03)   &    7.29      ( 0.02)   &    6.87      ( 0.02)   &    6.76      ( 0.01)  &                 II     &       C       \\
        31 &          G024.1902+00.4231 &  18h33m52.755s  & -7d34m43.65s   &    nan    (    nan)   &     nan       ( nan)    &    nan      (  nan)    &  11.94      ( 0.03)   &   11.29      ( 0.04)   &   10.66      ( 0.06)   &    9.98      ( 0.06)   &    8.31      ( 0.05)  &                 II     &       B     \\
        32 &          G024.0772+00.3604 &  18h33m53.598s  & -7d42m28.52s   &  13.78    (   0.04)   &   13.20       (0.06)    &  12.65      ( 0.06)    &  12.09      ( 0.08)   &   11.81      ( 0.06)   &   11.72      ( 0.12)   &     nan      (  nan)   &     nan      (  nan)  &                 II     &              \\
        33 &          G024.1165+00.3794 &  18h33m53.911s  & -7d39m51.58s   &  14.87    (   0.06)   &   14.11       (0.09)    &  13.60      ( 0.07)    &  13.24      ( 0.07)   &   12.88      ( 0.12)   &     nan      (  nan)   &     nan      (  nan)   &     nan      (  nan)  &                 II     &              \\
        34 &          G024.1193+00.3801 &  18h33m54.066s  & -7d39m41.49s   &  13.04    (   0.03)   &   12.63       (0.04)    &  12.32      ( 0.05)    &  12.34      ( 0.06)   &   12.03      ( 0.07)   &   11.77      ( 0.09)   &     nan      (  nan)   &     nan      (  nan)  &                 II     &              \\
        35 &          G024.0851+00.3603 &  18h33m54.500s  & -7d42m03.45s   &    nan    (    nan)   &   14.73       (0.08)    &  14.06      ( 0.09)    &  12.80      ( 0.08)   &   12.41      ( 0.11)   &     nan      (  nan)   &     nan      (  nan)   &     nan      (  nan)  &                 II     &              \\
        36 &          G024.1562+00.3942 &  18h33m55.162s  & -7d37m19.97s   &    nan    (    nan)   &     nan       ( nan)    &    nan      (  nan)    &  13.65      ( 0.07)   &   12.08      ( 0.06)   &   11.07      ( 0.06)   &     nan      (  nan)   &     nan      (  nan)  &                  I     &            \\
        37 &          G024.1569+00.3938 &  18h33m55.340s  & -7d37m18.47s   &    nan    (    nan)   &     nan       ( nan)    &    nan      (  nan)    &  13.76      ( 0.07)   &   12.05      ( 0.06)   &   10.82      ( 0.06)   &   10.12      ( 0.04)   &     nan      (  nan)  &                  I     &       B     \\
        38 &          G024.0843+00.3545 &  18h33m55.662s  & -7d42m15.79s   &    nan    (    nan)   &     nan       ( nan)    &    nan      (  nan)    &  13.30      ( 0.09)   &   12.49      ( 0.06)   &   11.71      ( 0.11)   &   11.17      ( 0.08)   &     nan      (  nan)  &                  I     &              \\
        39 &          G024.2214+00.4257 &  18h33m55.668s  & -7d32m59.47s   &  15.61    (   0.07)   &   14.25       (0.10)    &  13.49      ( 0.08)    &  12.68      ( 0.06)   &   12.52      ( 0.13)   &     nan      (  nan)   &     nan      (  nan)   &    7.50      ( 0.03)  &                  TD    &       A       \\
        40 &          G024.1622+00.3921 &  18h33m56.276s  & -7d37m04.33s   &    nan    (    nan)   &   12.79       (0.07)    &  10.53      ( 0.03)    &   8.02      ( 0.08)   &    6.97      ( 0.06)   &    6.27      ( 0.03)   &    5.56      ( 0.02)   &    3.57      ( 0.03)  &                  I     &       B     \\
        41 &          G024.2018+00.4077 &  18h33m57.350s  & -7d34m31.98s   &    nan    (    nan)   &     nan       ( nan)    &    nan      (  nan)    &  13.66      ( 0.07)   &   12.52      ( 0.09)   &   11.53      ( 0.11)   &     nan      (  nan)   &     nan      (  nan)  &                  I     &       A     \\
        42 &          G024.2060+00.4094 &  18h33m57.463s  & -7d34m15.70s   &    nan    (    nan)   &   13.66       (0.07)    &  13.37      ( 0.05)    &  13.03      ( 0.05)   &   12.40      ( 0.10)   &     nan      (  nan)   &     nan      (  nan)   &     nan      (  nan)  &                 II     &       A     \\
        43 &          G024.1613+00.3859 &  18h33m57.501s  & -7d37m17.68s   &    nan    (    nan)   &     nan       ( nan)    &    nan      (  nan)    &  11.51      ( 0.04)   &   10.33      ( 0.04)   &    9.64      ( 0.05)   &    9.08      ( 0.05)   &     nan      (  nan)  &                 II     &            \\
        44 &          G024.1619+00.3842 &  18h33m57.954s  & -7d37m18.64s   &    nan    (    nan)   &     nan       ( nan)    &  13.43      ( 0.05)    &  10.64      ( 0.03)   &    9.40      ( 0.03)   &    8.48      ( 0.03)   &    7.74      ( 0.02)   &    4.62      ( 0.02)  &                  I     &       B     \\
        45 &          G024.2459+00.4271 &  18h33m58.110s  & -7d31m39.03s   &  14.70    (   0.04)   &   13.95       (0.08)    &  13.51      ( 0.07)    &  13.12      ( 0.06)   &   12.84      ( 0.10)   &     nan      (  nan)   &     nan      (  nan)   &     nan      (  nan)  &                 II     &              \\
        46 &          G024.1232+00.3631 &  18h33m58.155s  & -7d39m57.12s   &  12.66    (   0.03)   &   12.29       (0.05)    &  11.95      ( 0.05)    &  11.85      ( 0.05)   &   11.55      ( 0.07)   &   11.55      ( 0.11)   &     nan      (  nan)   &     nan      (  nan)  &                 II     &              \\
        47 &          G024.1268+00.3648 &  18h33m58.194s  & -7d39m42.85s   &  14.78    (   0.07)   &   13.92       (0.05)    &  13.40      ( 0.06)    &  12.59      ( 0.06)   &   12.34      ( 0.09)   &     nan      (  nan)   &     nan      (  nan)   &     nan      (  nan)  &                 II     &              \\
        48 &          G024.1722+00.3867 &  18h33m58.556s  & -7d36m41.38s   &    nan    (    nan)   &     nan       ( nan)    &    nan      (  nan)    &  13.39      ( 0.07)   &   13.15      ( 0.13)   &     nan      (  nan)   &     nan      (  nan)   &    6.62      ( 0.05)  &                 TD     &       A     \\
        49 &          G024.2302+00.4136 &  18h33m59.243s  & -7d32m51.42s   &  13.78    (   0.04)   &   10.43       (0.04)    &   8.61      ( 0.03)    &   7.25      ( 0.04)   &    6.91      ( 0.04)   &    6.38      ( 0.02)   &    6.23      ( 0.02)   &    5.17      ( 0.03)  &                 II     &              \\
        50 &          G024.2098+00.3982 &  18h34m00.285s  & -7d34m22.28s   &    nan    (    nan)   &   12.70       (0.06)    &  10.79      ( 0.03)    &   8.67      ( 0.02)   &    7.94      ( 0.03)   &    7.30      ( 0.02)   &    6.43      ( 0.02)   &    4.10      ( 0.02)  &                 II     &       A     \\
        51 &          G024.1685+00.3761 &  18h34m00.434s  & -7d37m10.86s   &    nan    (    nan)   &   13.94       (0.08)    &  12.58      ( 0.06)    &  10.66      ( 0.04)   &    9.82      ( 0.04)   &    9.29      ( 0.03)   &    8.92      ( 0.02)   &    7.23      ( 0.03)  &                 II     &       B       \\
        52 &          G024.1168+00.3451 &  18h34m01.326s  & -7d40m47.36s   &  13.18    (   0.03)   &   12.49       (0.03)    &  12.16      ( 0.03)    &  11.91      ( 0.05)   &   11.76      ( 0.06)   &   11.42      ( 0.09)   &   10.81      ( 0.06)   &    7.86      ( 0.02)  &                 TD     &       A       \\
        53 &          G024.2221+00.3988 &  18h34m01.528s  & -7d33m42.03s   &    nan    (    nan)   &     nan       ( nan)    &    nan      (  nan)    &  12.39      ( 0.05)   &   12.09      ( 0.07)   &   11.89      ( 0.11)   &     nan      (  nan)   &    6.00      ( 0.04)  &                 TD     &       A       \\
        54 &          G024.0946+00.3326 &  18h34m01.532s  & -7d42m19.10s   &  14.70    (   0.05)   &   14.00       (0.06)    &  13.49      ( 0.07)    &  12.97      ( 0.07)   &   12.71      ( 0.08)   &     nan      (  nan)   &     nan      (  nan)   &     nan      (  nan)  &                 II     &              \\
        55 &          G024.1530+00.3596 &  18h34m02.245s  & -7d38m27.84s   &  12.66    (   0.03)   &   12.19       (0.04)    &  11.85      ( 0.04)    &  11.52      ( 0.08)   &   11.06      ( 0.07)   &   11.37      ( 0.15)   &     nan      (  nan)   &     nan      (  nan)  &                 II     &              \\
        56 &          G024.1220+00.3357 &  18h34m03.914s  & -7d40m46.57s   &    nan    (    nan)   &     nan       ( nan)    &  13.85      ( 0.10)    &  12.10      ( 0.08)   &   11.59      ( 0.12)   &   11.07      ( 0.16)   &   10.04      ( 0.14)   &     nan      (  nan)  &                 II     &              \\
        57 &          G024.1343+00.3389 &  18h34m04.593s  & -7d40m01.80s   &  13.88    (   0.03)   &   13.14       (0.04)    &  12.77      ( 0.04)    &  12.16      ( 0.08)   &   11.64      ( 0.10)   &     nan      (  nan)   &     nan      (  nan)   &     nan      (  nan)  &                 II     &              \\
        58 &          G024.1006+00.3190 &  18h34m05.118s  & -7d42m22.58s   &  15.25    (   0.05)   &   14.55       (0.07)    &  13.88      ( 0.08)    &  13.08      ( 0.08)   &   12.86      ( 0.10)   &     nan      (  nan)   &     nan      (  nan)   &     nan      (  nan)  &                 II     &              \\
        59 &          G024.1393+00.3376 &  18h34m05.448s  & -7d39m47.95s   &    nan    (    nan)   &   14.56       (0.09)    &  13.13      ( 0.06)    &  12.42      ( 0.09)   &   11.81      ( 0.05)   &   11.83      ( 0.18)   &     nan      (  nan)   &     nan      (  nan)  &                 II     &              \\
        60 &          G024.1441+00.3350 &  18h34m06.527s  & -7d39m37.15s   &    nan    (    nan)   &   13.71       (0.07)    &  13.42      ( 0.06)    &  13.17      ( 0.08)   &   12.90      ( 0.08)   &     nan      (  nan)   &     nan      (  nan)   &     nan      (  nan)  &                 II     &              \\
        61 &          G024.1581+00.3381 &  18h34m07.433s  & -7d38m47.10s   &    nan    (    nan)   &     nan       ( nan)    &    nan      (  nan)    &  11.76      ( 0.10)   &   10.91      ( 0.07)   &   10.36      ( 0.06)   &    9.12      ( 0.03)   &     nan      (  nan)  &                 II     &              \\
        62 &          G024.1740+00.3446 &  18h34m07.803s  & -7d37m45.61s   &    nan    (    nan)   &     nan       ( nan)    &    nan      (  nan)    &  13.04      ( 0.10)   &   12.35      ( 0.08)   &   11.60      ( 0.10)   &   10.77      ( 0.06)   &     nan      (  nan)  &                 II     &              \\
        63 &          G024.1981+00.3570 &  18h34m07.830s  & -7d36m07.78s   &  14.75    (   0.04)   &   14.17       (0.06)    &  13.94      ( 0.08)    &  13.49      ( 0.08)   &   13.20      ( 0.12)   &     nan      (  nan)   &     nan      (  nan)   &     nan      (  nan)  &                 II     &              \\

\enddata
\tablenotetext{Notes}{$\dag$ YSOs associated with the core are confined over 10 $\sigma$ contours of \13co\ integrated intensity;
* Protostars classified as sources with PAH-contaminated apertures were re-identified visually from the bright 24 $\mu$m emission.}
\end{deluxetable*}
\clearpage
\end{landscape}
\end{document}